\documentclass[useAMS,usenatbib]{mn2e}
\usepackage{graphicx}
\usepackage{times}
\usepackage{natbib}
\usepackage{rotating}
\usepackage{color}
\usepackage{amsmath}
\usepackage{amssymb}

\renewcommand{\vec}[1]{\bmath{#1}}
\newcommand{\be}{\begin{equation}}
\newcommand{\ee}{\end{equation}}
\newcommand{\ba}{\begin{eqnarray}}
\newcommand{\ea}{\end{eqnarray}}
\newcommand{\brr}{\begin{array}}
\newcommand{\err}{\end{array}}
\newcommand{\bc}{\begin{center}}
\newcommand{\ec}{\end{center}}

\newcommand{\mincir}{\raise
  -2.truept\hbox{\rlap{\hbox{$\sim$}}\raise5.truept \hbox{$<$}\ }}
\newcommand{\magcir}{\raise
  -2.truept\hbox{\rlap{\hbox{$\sim$}}\raise5.truept \hbox{$>$}\ }}
\newcommand{\siml}{\raise
  -2.truept\hbox{\rlap{\hbox{$\sim$}}\raise5.truept \hbox{$<$}\ }}
\newcommand{\simg}{\raise
  -2.truept\hbox{\rlap{\hbox{$\sim$}}\raise5.truept \hbox{$>$}\ }}

\title[SZ effects in the Magneticum Pathfinder Simulation]{
SZ effects in the Magneticum Pathfinder Simulation: Comparison with
the Planck, SPT, and ACT results}

\author[K. Dolag, E. Komatsu, R. Sunyaev]
{  K.~Dolag$^{1,2}$\thanks{E-mail: dolag@usm.uni-muenchen.de}, E.~Komatsu$^{2,3}$ and R. Sunyaev$^{2,4}$ \\
$^1$ University Observatory Munich, Scheinerstr. 1, 81679 Munich, Germany\\
$^2$ Max-Planck-Institut f\"ur Astrophysik, Karl-Schwarzschild Strasse
  1, 85748 Garching, Germany\\
$^3$ Kavli Institute for the Physics and
Mathematics of the Universe (Kavli IPMU, WPI), Todai Institutes for Advanced Study, the
University of Tokyo, \\Kashiwa 277-8583, Japan \\
$^4$ Space Research Institute (IKI), Russian Academy of Sciences, Profsoyuznaya
str. 84/32, Moscow, 117997 Russia
}
 
\begin{document}

\date{Accepted ???. Received ???; in original form ???}

\pagerange{\pageref{firstpage}--\pageref{lastpage}} \pubyear{0000}

\maketitle

\label{firstpage}

\begin{abstract}
 We calculate the one-point probability density distribution
 functions (PDF) and the power spectra of the thermal and kinetic
 Sunyaev-Zeldovich (tSZ and kSZ) effects and the mean Compton $Y$ parameter using the {\it{}Magneticum
 Pathfinder} simulations, state-of-the-art cosmological hydrodynamical
 simulations of a large cosmological volume of
 $(896~\mathrm{Mpc}/h)^3$. These simulations follow in detail the
 thermal and chemical evolution of the intracluster medium as well as
 the evolution of super-massive black holes and their associated feedback
 processes. We construct full-sky maps of tSZ and kSZ from
 the light-cones out to $z=0.17$, and one realisation of
 $8^\circ.8\times{}8^\circ.8$ deep light-cone out to $z=5.2$. The local
 universe at $z<0.027$ is simulated by a constrained realisation. The
 tail of the one-point PDF of tSZ from the deep light-cone follows a
 power-law shape with an index of $-3.2$. Once convolved with the
 effective beam of Planck, it agrees with the PDF measured by
 Planck. The predicted tSZ power spectrum agrees with that of the
 Planck data at all multipoles up to $l\approx{}1000$, once the
 calculations are scaled to the Planck 2015 cosmological parameters with
 $\Omega_m=0.308$ and $\sigma_8=0.8149$. Consistent with the results in
 the literature, however, we continue to find the tSZ power spectrum at
 $l=3000$ that is significantly larger than that estimated from the
 high-resolution ground-based data.
 The simulation predicts the mean fluctuating Compton $Y$ value of
 $\bar{}Y=1.18\times10^{-6}$ for $\Omega_m=0.272$ and $\sigma_8=0.809$.
 Nearly half ($\approx{}5\times10^{-7}$) of the
 signal comes from halos below a virial mass of $10^{13}~M_\odot/h$.
 Scaling this to the Planck 2015 parameters, we find $\bar Y=1.57\times{}10^{-6}$.
\end{abstract}



\section{Introduction} \label{sec:intro}

One of the last unsolved problems in cosmology with  the cosmic
microwave background (CMB) is connected
with unknown levels of the spectral distortion of the black-body
spectrum of CMB. Two decades ago the Far
Infrared Absolute Spectrophotometer (FIRAS) on board NASA's Cosmic
Background Explorer (COBE) provided the first limits \citep{1996ApJ...473..576F} on the so-called
$y$- and $\mu$-type spectral distortions, which originate from energy releases
in the present and early universe
\citep{1970CoASP...2...66S,1980ARA&A..18..537S}. An enormous progress in
technology of cryogenics and detectors of millimetre and sub-millimetre
radiation enables us now to reach sensitivity which is 100 or even 1000
times better than that of FIRAS. A recently proposed PIXIE instrument is
an example \citep{kogut/etal:2014}.

While measuring the absolute monopole of the spectral distortions
requires an absolute spectrometer such as FIRAS and PIXIE, the {\it
fluctuating} component does not. \citet{2015arXiv150500781K}
have recently demonstrated that the Planck data \citep{2015arXiv150201596P} permit to
limit the mean of the fluctuating part of the $y$-type distortion to be
$\bar Y<2.2 \times 10^{-6}$ (where $Y$ is the so-called Compton $Y$
parameter), thereby reducing the observational upper bound by a
factor of seven compared to the original COBE/FIRAS limit.\footnote{However, we
should not forget that the COBE/FIRAS limit applies to the sum of
the uniform and the fluctuating parts, whereas the Planck limit applies
only to the latter.}

The $y$-type spectral distortions do not carry information about
redshifts at which they were produced. Therefore, it will be a challenge
to distinguish between the primordial $y$-distortions originated from
the pre-recombination era and those from a low-redshift universe induced by the large-scale
structures. Nevertheless, in the standard thermal history of
the universe, the latter component is expected
to dominate by a couple of orders of magnitude\footnote{In principle,
knowing the precise expectation for the low-redshift tSZ contribution
would make it possible to estimate the detectable level of the primordial $y$-type
distortions from the pre-recombination era.
In the absence of primordial non-Gaussianity, we do not expect any
angular dependence of the $y$-type signal before recombination.
Therefore, characterizing the angular dependence of the
fluctuating component of the $y$-signal
from the large-scale structures might open a possibility to
separate that contribution from the primordial monopole (via, e.g.,
cross-correlation of tSZ with other tracers of the large-scale
structure). If successful, we can retrieve information on the energy
release at redshifts of $1000 \lesssim z \lesssim 20000$.}; namely, the dominant
component of the mean $Y$ is expected to come from the fluctuating
component of the thermal Sunyaev-Zeldovich (tSZ) effect, i.e., 
inverse Compton scattering of low energy CMB photons on hot non-relativistic electrons in groups and clusters of
galaxies \citep[][]{1970Ap&SS...7....3S}. Therefore, the
new Planck bound provides a limit on the total thermal energy content
of hot gas in the universe. Indeed, this 
bound is approaching the predicted value of the mean $Y$ calculated by
the previous generations of cosmological hydrodynamical simulations; for
example, \citet{refregier/etal:2000} find $\bar Y=1.67\times 10^{-6}$ for a flat
$\Lambda$ Cold Dark Matter (CDM) model with $\Omega_m=0.37$ and
$\sigma_8=0.8$ \citep[also see][]{2000MNRAS.317...37D,seljak/etal:2001,2004MNRAS.355..451Z}.
It is therefore timely to revisit this calculation with
much improved, state-of-the-art simulations.

The Planck satellite has measured the power spectrum of fluctuations of
tSZ on large angular scales ($l\lesssim 1000$). The tSZ power
spectrum is very sensitive to the amplitude of matter density
fluctuations \citep{komatsu/kitayama:1999,komatsu/seljak:2002}. The power spectrum on large angular scales is particularly
a powerful probe of the amplitude of fluctuations, as it is less
sensitive to astrophysics of the core of galaxy clusters than that
at $l\approx 3000$ \citep{komatsu/kitayama:1999,2014MNRAS.440.3645M}.

Cosmological hydrodynamical simulations have proven
 essential in interpreting the observational data on the statistics of
 tSZ such
 as the power spectrum and the one-point probability density
 distribution function (PDF). The simulations have steadily improved
 over the past two decades, in terms of the numerical methods, the
 volume, the mass and
 spatial resolution, and implementation
 of baryonic physics such as cooling, heating, chemistry, and feedback processes
 \citep{persi/etal:1995,refregier/etal:2000,2000MNRAS.317...37D,seljak/etal:2001,2001ApJ...549..681S,dasilva2001,2002PhRvD..66d3002R,white2002,2002ApJ...577..555Z,dolag2005,2007ApJ...671...27H,2007MNRAS.378.1259R,2010ApJ...725.1452S,2010ApJ...725...91B,2012ApJ...758...75B,2013MNRAS.429.1564M,2014MNRAS.440.3645M}. 

The recent progress in performing larger and more ``complete'' cosmological
 hydrodynamical simulations, which follow baryons in different phases
 such as in Warm-Hot Intergalactic Medium (WHIM,\citet{1999ApJ...514....1C}), Intracluster Medium
 (ICM), stars and super massive black holes and their related feedback,
 allows now for a direct and detailed comparison of the predicted tSZ
 signals with observations. Such comparison may shed light on
 the recently reported tensions between cosmological parameters inferred
 from the primary CMB with the ones inferred from tSZ
 \citep{2015arXiv150201596P,2015arXiv150201597P}. 

Scattering of CMB photons off electrons moving with a non-zero
line-of-sight bulk velocity with respect to the frame of coordinates where the CMB is isotropic yields temperature
fluctuations by the Doppler shift of light, and this is known as the kinetic Sunyaev-Zeldovich (kSZ)
effect \citep{1970Ap&SS...7....3S,1980MNRAS.190..413S}. There are two contributions to the
kSZ signal: the reionisation contribution from $z\gtrsim 6$, and the
post-reionisation contribution. The calculation of the former requires
detailed reionisation simulations including radiative transfer
 and is beyond
the scope of this paper. The calculation of the latter is in principle
simpler than the former. Precisely characterising the post-reionisation
kSZ is important, as we must subtract this contribution from the total
kSZ to extract the reionisation contribution which, in turn, can
constrain the physics of reionisation \citep[see][and references
therein]{2012ApJ...756...65Z,2013ApJ...769...93P}. Roughly speaking, the post-reionisation
kSZ power spectrum is twice as large as that from reionisation; thus,
ten per cent uncertainty in the post-reionisation contribution results in
twenty per cent uncertainty in the reionisation contribution \citep{2015arXiv150605177P}.

In a fully ionised universe, the kSZ effect is
given by the line-of-sight integration of the radial momentum field, i.e., a large-scale velocity field
modulated by a small-scale density fluctuations of
electrons \citep[see][and references therein]{2015arXiv150605177P}. Therefore, the calculation of the post-reionisation kSZ requires a
simulation whose box size is large enough to capture the long-wavelength
velocity field, and the spatial resolution high enough to resolve
non-linear structures of baryons responsible for scattering of CMB
photons. The hydrodynamical simulations of the post-reionisation kSZ have also improved  over the past decade
\citep{2001ApJ...549..681S,2001MNRAS.326..155D,dasilva2001,white2002,2002ApJ...577..555Z,2004MNRAS.347.1224Z,2007MNRAS.378.1259R,2012ApJ...756...15S,2013MNRAS.432.1600D}. The
measurements of the post-reionisation kSZ are improving rapidly as well
\citep{2012PhRvL.109d1101H,2014MNRAS.443.2311L}. 

In this paper, we shall push the simulation frontier on the investigation
of tSZ and kSZ further.
The paper is organized as follows. Section 2 reviews the simulation method
and the light-cone generation. Section 3 shows detailed comparisons
of the simulation results with the observational data on the Coma
cluster, and the one-point PDF and power spectrum of tSZ. We also show
the simulation predictions for kSZ toward Coma, and the one-point PDF
and power spectrum of kSZ.
In section 4,
we study how the mean
Compton $Y$ signal builds up over cosmic time. We summarise our findings
in section 5.

\section{Simulations}

We construct sky maps of tSZ and kSZ using two sets of
simulations: the ``local universe'' simulation (Sec.~\ref{sec:local})
based on a constrained realisation of the local universe at $z<0.027$,
and the {\it Magneticum Pathfinder} simulation
(Sec.~\ref{sec:magneticum}). Combining these simulations, we construct
full-sky maps of tSZ and kSZ out to $z=0.17$, and one
realisation of $8^\circ.8\times 8^\circ.8$ deep light-cone out to $z=5.2$.

Both simulations are based on the parallel cosmological Tree
Particle-Mesh (PM) Smoothed-particle Hydrodynamics (SPH) code {\small
P-GADGET3} (\citealp{Springel05gad}). The code uses an
entropy-conserving formulation of SPH \citep{2002MNRAS.333..649S} and
follows the gas using a low-viscosity SPH scheme to properly track
turbulence \citep{2005MNRAS.364..753D}. Based on \citet{2004ApJ...606L..97D},
it also follows thermal conduction at 1/20th of the classical Spitzer
value \citep{1962pfig.book.....S}. It also allows a treatment of
radiative cooling, heating from a uniform time-dependent ultraviolet
background, and star formation with the associated feedback processes. 

Radiative cooling rates are computed by following the same procedure
presented by \citet{Wiersma09}. We account
for the presence of the CMB and
the ultraviolet (UV)/X-ray background radiation from quasars and
galaxies, as computed by \citet{Haardt01}. The contributions
to cooling from each one of 11 elements (H, He, C, N, O,
Ne, Mg, Si, S, Ca, Fe) have been pre-computed using the publicly
available CLOUDY photo-ionization code \citep{Ferland98} for an
optically thin gas in (photo-)ionization equilibrium. 

We model the interstellar medium (ISM) by using a sub-resolution model
for the multiphase ISM of \citet{Springel03}.
In this model, the ISM
is treated as a two-phase medium, in which clouds of cold gas form by
cooling of hot gas, and are embedded in the hot gas phase assuming
pressure equilibrium whenever gas particles are above a given threshold 
density. The hot gas within the multiphase model is heated by supernovae
and can evaporate the cold clouds. A certain fraction of massive stars (10 per
cent) is assumed to explode as supernovae type II (SNII). The released
energy by SNII ($10^{51}$~erg) triggers galactic winds
with a mass loading rate proportional to the star formation rate
(SFR) with a resulting wind velocity of $v_{\mathrm{wind}} = 350$~km/s.

We include a detailed model of chemical evolution 
according to \citet{Tornatore07}. Metals are produced by SNII, by supernovae
type Ia (SNIa) and by intermediate and low-mass stars in the asymptotic giant
branch (AGB). Metals and energy are released by stars of different masses, by
properly accounting for mass-dependent life-times (with a lifetime function
according to \citealp{Padovani93}), the metallicity-dependent stellar
yields by \citet{Woosley95} for SNII, the yields by \citet{vandenHoek97} for
AGB stars, and the yields by \citet{Thielemann03} for SNIa. Stars of different
masses are initially distributed according to a Chabrier initial mass function 
\citep[IMF;][]{Chabrier03}.

Most importantly, our simulations include prescriptions for
the growth of black holes and the feedback from active galactic nuclei (AGN)
based on the model of \citet{Springel05a} and
\citet{DiMatteo05} with the same modifications as in
\citet{Fabjan10} and some new, minor changes as described below.

\begin{figure*}
\begin{center}
\includegraphics[width=0.87\textwidth]{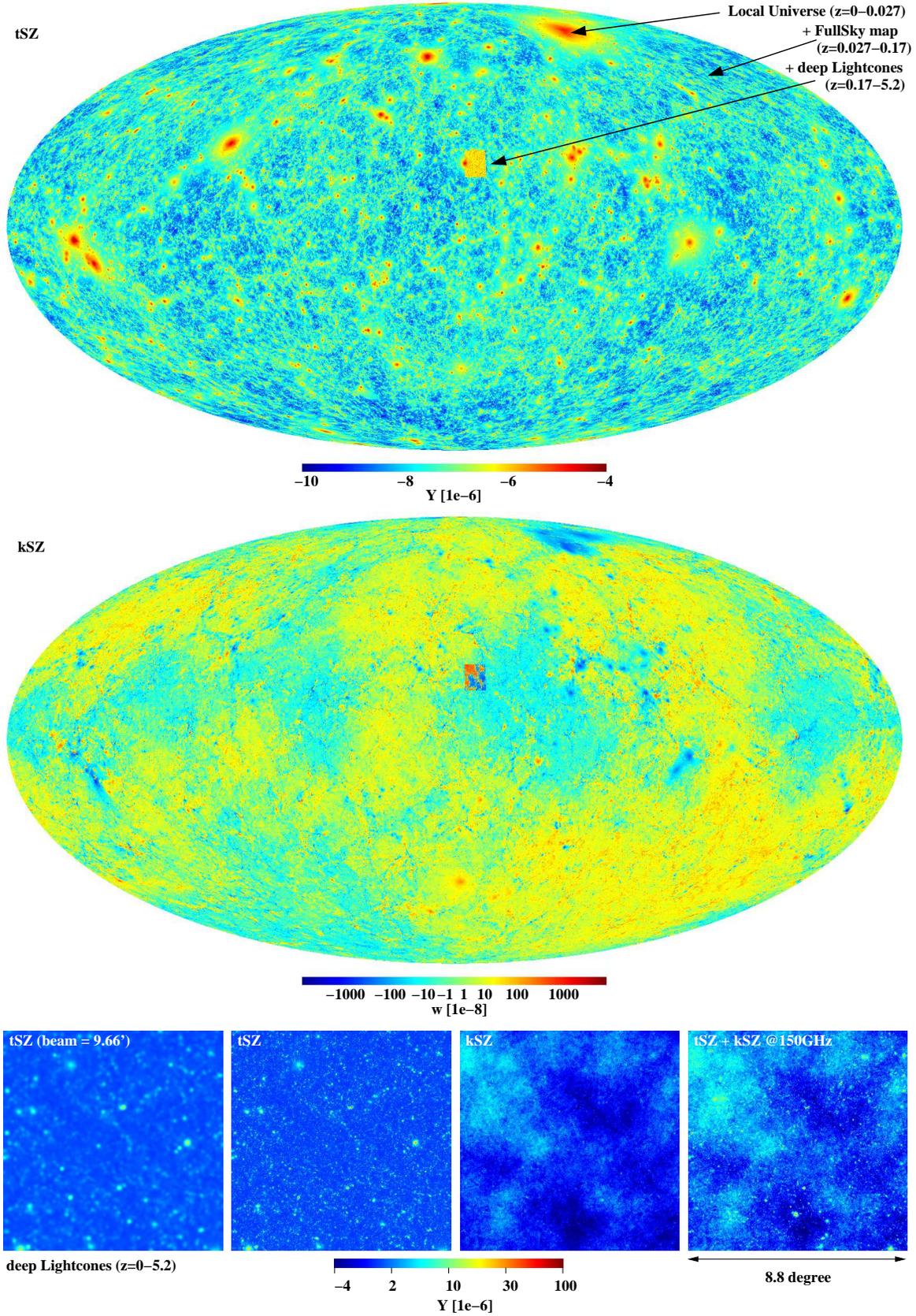}
\end{center}
\caption{Full-sky maps of the Compton $Y$ parameter
 (Eq.~\ref{eq:y_theta}; top) and the kSZ effect
 (Eq.~\ref{eq:w_theta}; middle) 
 obtained from the local universe simulation ($z<0.027$) combined with the full sky maps of the Magneticum Pathfinder simulation ($0.027<z<0.17$).
 We also show $8^\circ.8\times8^\circ.8$ maps from the deep light-cone of the Magneticum Pathfinder
 simulation restricted to $0.17<z<5.2$. The lower panels show, from the left to
 right: the Compton $Y$ from the deep light-cone ($0<z<5.2$) smoothed with a 9.66 arcmin
 FWHM Gaussian beam; the Compton $Y$ at the native resolution of HEALPix with
 nside=2048; the kSZ at the native resolution, and the sum of the two at
 150~GHz (Eq.~\ref{eq:yeff}).}
\label{fig:map}
\end{figure*}

The accretion onto black holes and the associated
feedback adopts a sub-resolution model. Black holes are represented by 
collisionless ``sink particles'', which can grow in mass by either accreting gas
from their environments, or merging with other black holes. 
The gas accretion rate, $\dot{M}_\bullet$, is estimated by the
Bondi-Hoyle-Lyttleton approximation \citep{Hoyle39, Bondi44,
  Bondi52}: 
\begin{equation}\label{Bondi}
\dot{M}_\bullet = \frac{4 \pi G^2 M_\bullet^2 f_{\rm boost} \rho}{(c_s^2 + v^2)^{3/2}},
\end{equation}
where $\rho$ and $c_s$ are the density and the sound speed of the
surrounding (ISM) gas, respectively, $f_{\rm boost}$ is a boost factor
for the density which typically is set to $100$ and $v$ is the velocity of the
black hole relative to the surrounding gas. The black hole accretion
is always limited to the Eddington rate, i.e., the maximum possible accretion
for balance between inwards-directed gravitational force and outwards-directed radiation pressure: $\dot{M}_\bullet = \min(\dot{M}_\bullet,
\dot{M}_{\mathrm{edd}})$. Since the detailed accretion flows onto
the black holes are unresolved, we can only capture black hole growth
due to the larger scale, resolved gas distribution.

Once the accretion rate is computed for each BH particle,
the mass continuously grows. To model the loss of this gas from the gas
particles, a stochastic criterion is used to select the surrounding gas
particles that are accreted. Unlike in \citet{Springel05a}, in which a
selected gas particle contributes to accretion with all its mass, we
include the possibility for a gas particle to accrete only with a
fraction (1/4) of its original mass. In this way,
each gas particle can contribute with up to four `generations' of BH 
accretion events, thus providing a more continuous description of the
accretion process.

As for the feedback, the radiated luminosity,
$L_{\mathrm{r}}$, is related to the black hole accretion rate by
\begin{equation}
L_{\mathrm{r}} = \epsilon_{\mathrm{r}} \dot{M}_\bullet c^2,
\end{equation} 
where $\epsilon_{\mathrm{r}}$ is the radiative efficiency, for which
we adopt a fixed value of 0.1. This value is assumed typically for a radiatively
efficient accretion disk onto a non-rapidly spinning black hole \citep{Shakura73,Springel05gad, DiMatteo05}. We assume that a fraction $\epsilon_{\mathrm{f}}$
of the radiated energy is thermally coupled to the surrounding gas; thus,
$\dot{E}_{\mathrm{f}} = \epsilon_{\mathrm{r}}
\epsilon_{\mathrm{f}} \dot{M}_\bullet c^2$ is the rate of the energy
feedback. $\epsilon_{\mathrm{f}}$ is a free parameter and typically set
to $0.1$. The energy is distributed to the surrounding
gas particles with weights similar to SPH. In addition, we incorporate the
feedback prescription of \citet{Fabjan10}; namely, we account for a
transition from a quasar- to a radio-mode feedback (see also
\citealp{Sijacki07}) whenever the accretion rate falls below an
Eddington-ratio of $f_{\mathrm{edd}} \equiv \dot{M}_\bullet/
\dot{M}_{\mathrm{edd}} < 10^{-2}$. During the radio-mode feedback we
assume a 4 times larger feedback efficiency than in the quasar mode.
This way, we attempt to account for massive black holes that are 
radiatively inefficient (having low accretion rates) but are
efficient in heating the ICM by inflating hot bubbles in 
correspondence of the termination of AGN jets. 
In contrast to \citet{Springel05a}, we modify the mass growth of the
BH by taking into account the feedback, e.g.,
$ \Delta M_\bullet = (1-\epsilon_{r})\dot{M}_\bullet \Delta
t$.
We introduced some more technical modifications of the
original implementation, for which readers can find details in 
\citet{2014MNRAS.442.2304H}, where we also demonstrate that the 
bulk properties of the AGN population within the simulation
are quite similar to the observed AGN properties.

\begin{figure*}
\includegraphics[width=1.0\textwidth]{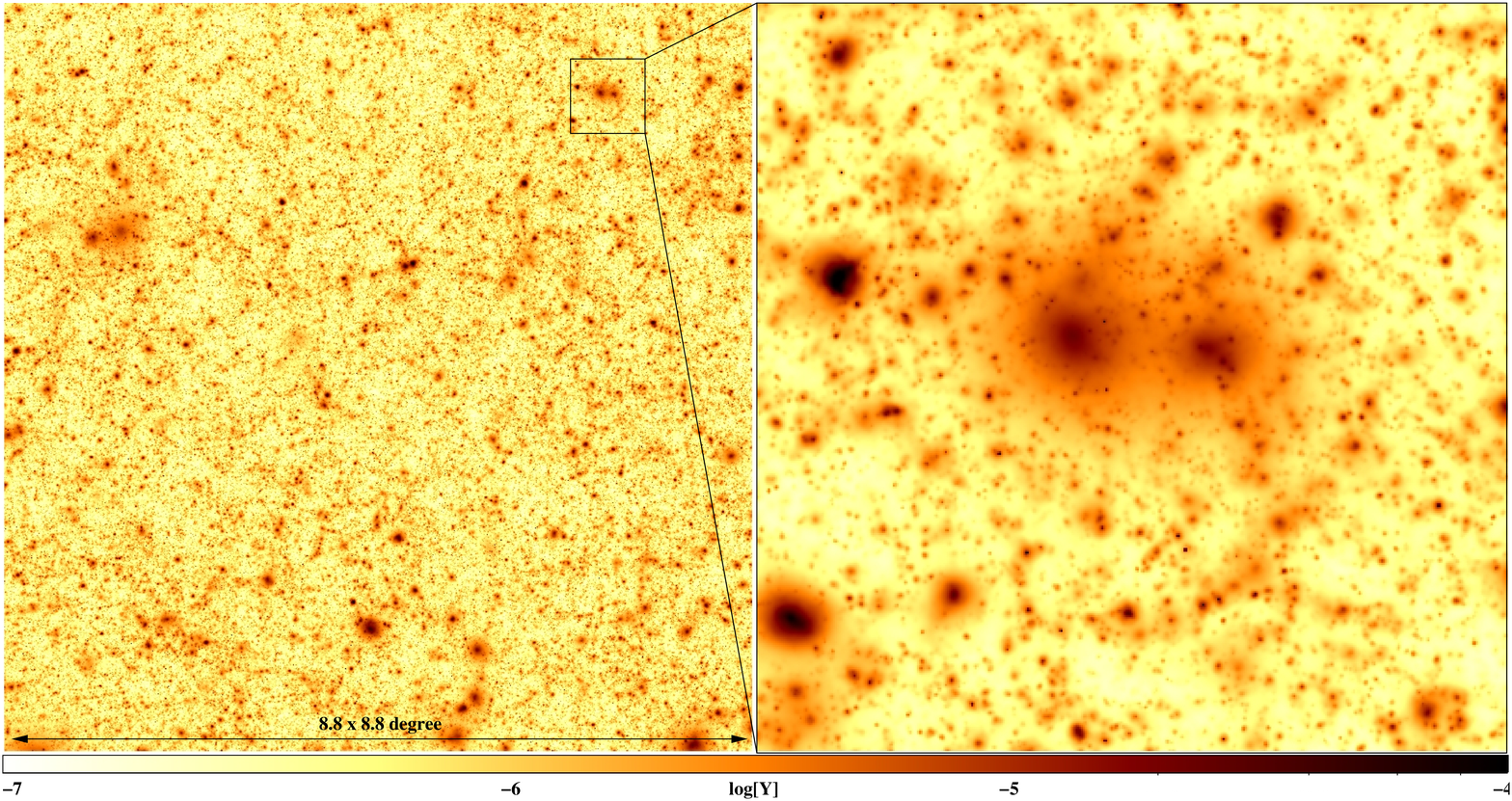}
\caption{Compton $Y$ from the deep light-cone ($0<z<5.2$).
The right panel shows a zoom onto a region
containing several rich clusters at various redshifts.}
\label{fig:cone}
\end{figure*}

\subsection{Local Universe Simulation}
\label{sec:local}
The local universe simulation uses the final output of a
cosmological hydrodynamical simulation of a constrained realisation of
the local universe based on the nearly full-sky IRAS 1.2-Jy galaxy survey
data \citep{fisher/etal:1994,fisher/etal:1995}.
The initial conditions are similar to those adopted by
\citet{Mathis:2002} in their study of structure formation in the local
universe. The galaxy distribution in
the IRAS 1.2-Jy galaxy survey is first smoothed by a Gaussian with a width of 7
Mpc and then linearly evolved back in time up to $z=50$ following the
method of \cite{Kolatt:1996}. The resulting field is then used
as a constraint on phases of random Gaussian fields
\citep{Hoffman1991} for initialising the simulation. 

The volume constrained by the observational data covers a sphere of
radius $\approx 80~{\rm Mpc}/h$ centered on the Milky Way. This region is sampled with
more than 50 million high-resolution dark matter particles and is
embedded in a periodic box of $\approx 240~{\rm Mpc}/h$ on a side. The
region outside the constrained volume is filled with nearly 7 million
low-resolution dark matter particles, allowing a good coverage of
long-range gravitational tidal forces.
The gravitational force resolution (i.e., the softening length) of
the simulation has been fixed to be $7~{\rm kpc}/h$
(Plummer-equivalent), fixed in physical units from $z=0$ to $z=2$ and
then stays constant in the corresponding co-moving units (e.g. $21~{\rm
kpc}/h$) at higher redshifts.

Unlike in the original simulation made by \citet{Mathis:2002}, where only the
dark matter component is present, here we follow also the gas and
stellar components. Therefore, we extend the initial conditions by
splitting the original high-resolution dark matter particles into gas and dark
matter particles having masses of $m_\mathrm{gas} \approx 0.48 \times 10^9\; M_\odot/h$
and $m_\mathrm{dm} \approx3.1 \times 10^9\; M_\odot/h$, respectively; 
this corresponds to a cosmological baryon
fraction of 13 per cent. The total number of particles within the simulation
is slightly more than 108 million, and the most massive cluster is
resolved by almost one million particles.

In this simulation, we
assume a flat $\Lambda$CDM model with a present matter density
parameter of $\Omega_m=0.3$, a Hubble constant of $H_0=100~h$ km/s/Mpc
with $h=0.7$, and
an r.m.s. density fluctuation of $\sigma_8=0.9$.

\subsection{Magneticum Pathfinder Simulation}
\label{sec:magneticum}

The {\it Magneticum Pathfinder}\footnote{www.magneticum.org} simulation follows a large
$(896~\mathrm{Mpc}/h)^3$ box simulated using $2\times1526^3$
particles and adapting a WMAP7 \citep{Komatsu11} flat $\Lambda$CDM
cosmology with $\sigma_8 = 0.809$, $h = 0.704$, $\Omega_m = 0.272$, 
$\Omega_b = 0.0456$, and an initial slope for the power spectrum of $n_s
= 0.963$. Dark matter particles have a 
mass of $m_\mathrm{DM} = 1.3 \times 10^{10}~M_\odot/h$, gas particles 
have $m_\mathrm{gas} \approx 2.6 \times 10^9~M_\odot/h$ 
depending on their enrichment history, and stellar particles have
$m_\mathrm{stars} \approx 7.5 \times 10^8~M_\odot/h$ depending on the state
of the underlying stellar population. For gas and dark matter,
the gravitational softening length is set to $10~{\rm kpc}/h$
(Plummer-equivalent), fixed in physical units from $z=0$ to $z=2$ and
then stays constant in the corresponding co-moving units
(e.g. $30~{\rm kpc}/h$) at higher redshifts. For star particles it is accordingly half
the values (e.g. $5~{\rm kpc}/h$ at $z=0$). In this simulation, one gas particle can form up to four
stellar particles.

\begin{figure*}
  \includegraphics[width=0.99\textwidth]{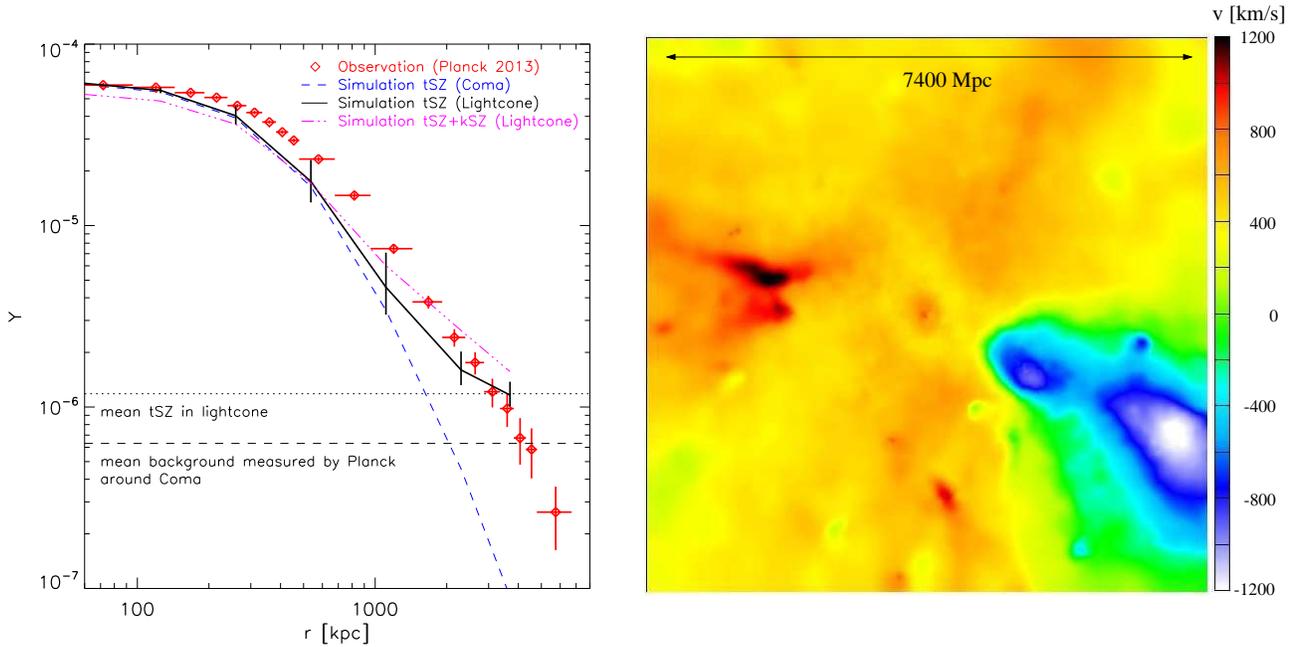}
\caption{Compton $Y$ profile of the Coma cluster (left panel). The
 symbols show the Planck data taken from
 \citet{2013A&A...554A.140P}. These data points have an
 estimate of the mean background (horizontal dashed line) subtracted by
 the Planck team. The blue line shows the Compton $Y$ toward the
 counterpart of the Coma cluster in the local universe simulation, while
 the thick black solid line shows the Coma cluster embedded in the deep
 light-cone of the Magneticum Pathfinder simulation, to show the effect
 of the contribution of gas outside the Coma cluster. 
 All simulated maps are smoothed with a Gaussian beam of 10 arcmin.
 The error bars on the thick black solid line show the
 25\% and 75\% percentile of the Compton $Y$ distribution in each radial
 bin, averaged over 9 different maps obtained by placing Coma at
 different positions within the light-cone. The horizontal dotted
 line shows the mean Compton $Y$ of the deep light-cone, $\bar
 Y=1.18\times 10^{-6}$. The magenta line shows the
 sum of tSZ and kSZ (Eq.~\ref{eq:yeff}) at 150~GHz. 
 In the local universe simulation, Coma is moving away from us, thus
 giving a negative kSZ near the center. On the other hand, a positive
 kSZ in the outskirt is due to a massive, gas rich sub-structure moving
 toward us, as shown in the mass-weighted velocity map of Coma in the
 local universe simulation (right panel).
} 
\label{fig:coma}
\end{figure*}

\subsection{Map Making}
\label{sec:map}
We construct maps from the simulations using {\small SMAC}
\citep{dolag2005}, integrating both the tSZ and kSZ signals through our
hydrodynamical simulations. 

The tSZ signal in each pixel at a sky position
$\theta$ is characterised by the Compton $Y$ parameter defined by
\citep{1972CoASP...4..173S}
\begin{equation}
Y_\mathrm{tSZ}(\vec{\theta})= {{k_B\sigma_{T}}\over{m_e c^2}} \int \mathrm{d} l
\ n_e(\vec{\theta},l)\ T(\vec{\theta},l)\ ,
\label{eq:y_theta}
\end{equation}
where $n_e$ and $T$ are the three-dimensional number density and
temperature of thermal electrons,
respectively, and $k_B$, $\sigma_T$, $m_e$, and $c$ are the Boltzmann
constant, the Thomson scattering cross section, the electron mass, and
the speed of light, respectively. The kSZ signal is obtained by
\citep{1970Ap&SS...7....3S,1980MNRAS.190..413S}
\begin{equation}
w_\mathrm{kSZ}(\vec{\theta})\equiv
\frac{\delta T_{\rm kSZ}}{T_{\rm cmb}}=
 - \frac{\sigma_{_T}}{c} \int \mathrm{d} l
\ n_e(\vec{\theta},l)\ v_r(\vec{\theta},l)\ ,
\label{eq:w_theta}
\end{equation}
where $T_{\rm cmb}=2.725~{\rm K}$ and $v_r$ is the radial component of the
peculiar bulk velocity, defined such that $v_r>0$ for gas moving away from us.

As the tSZ and kSZ have different dependence on the observed
frequencies, we use the following weighted sum of the two when showing
the combined signal:
\begin{equation}
  Y_\mathrm{eff}^\mathrm{frq} \equiv \frac{g(x) Y_\mathrm{tSZ} -
   w_\mathrm{kSZ}}{g(x)}\ ,
  \label{eq:yeff}
\end{equation}
with
\begin{equation}
g(x) = \frac{x (e^x+1)}{(e^x-1)}-4\ ,
\end{equation}
derived by \citet{1969Ap&SS...4..301Z}, where $x = \mathrm{frq} [GHz] /
56.8$. The temperature anisotropy due to tSZ is given by $\delta T_{\rm
tSZ}/T_{\rm cmb}=g(x)Y$, where we ignore relativistic corrections.

We produced full-sky maps of the Compton $Y$ and kSZ in the
HEALPix \citep{healpix} format with Nside=2048 from the local universe ($0<z<0.027$) and from
the Magneticum Pathfinder simulation covering $0.027<z<0.17$. We also produced one realization
of a deep, $8^\circ.8\times8^\circ.8$ light-cone covering
$0<z<5.2$. When we combine the deep light-cone with the full-sky maps,
we use only the relevant parts, $0.027<z<5.2$ or $0.17<z<5.2$, of the deep
light-cone.

In figure~\ref{fig:map}, we show the full-sky maps of the Compton $Y$ (top panel) and
kSZ (middle) where the local universe simulation is combined with the Magneticum Pathfinder
simulation. We also show $0.17<z<5.2$ of the
deep light-cone (as indicated by the arrow) to give the impression
of the additional SZ signals covered by the deep light-cone. The lower
panels show the Compton $Y$ from the deep light-cone ($0<z<5.2$) smoothed with a 9.66 arcmin
FWHM Gaussian beam as well as the  native resolution and the same with the kSZ signal
added at 150~GHz (Eq.~\ref{eq:yeff}). The deep light-cone contains about
8000 galaxy
clusters and groups with virial masses above $10^{13.5}~M_\odot/h$, contributing to the most
prominent structures visible in the zoom onto a sub-part of the Compton $Y$ map shown in the right
panel of figure~\ref{fig:cone}.

\section{Comparison with the Planck data} \label{sec:planck}
\subsection{Coma}

The thermal electron pressure determines the (non-relativistic) tSZ
effect. Azimuthally-averaged radial profiles of thermal pressure in
galaxy clusters identified in our hydrodynamical simulations follow the
so-called ``universe pressure profile'' \citep{2010A&A...517A..92A}. The
stacked pressure profiles of galaxy-cluster-size halos in our simulation
are in good agreement with the stacked pressure profiles inferred from
the tSZ data on galaxy clusters detected by Planck \citep{2013A&A...550A.131P} and South Pole
Telescope (SPT) \citep{2014ApJ...794...67M}.
However, such
comparisons allow only a statistical comparison of the averaged profiles.
	
Local, well-resolved galaxy clusters enable a more detailed
object-by-object comparison. 
Here we make use of the fact that
the constrained simulation of the local universe allows to cross-identify objects in the
simulations with the real-world counterparts. 
Despite the relatively low spacial resolution (e.g., $> 5$~Mpc) of the
constraints used for initialising the simulation \citep[see][for details]{Mathis:2002},
local galaxy clusters like the Coma cluster have a remarkably similar counterpart in the
simulation.

\begin{figure*}
\includegraphics[width=0.49\textwidth]{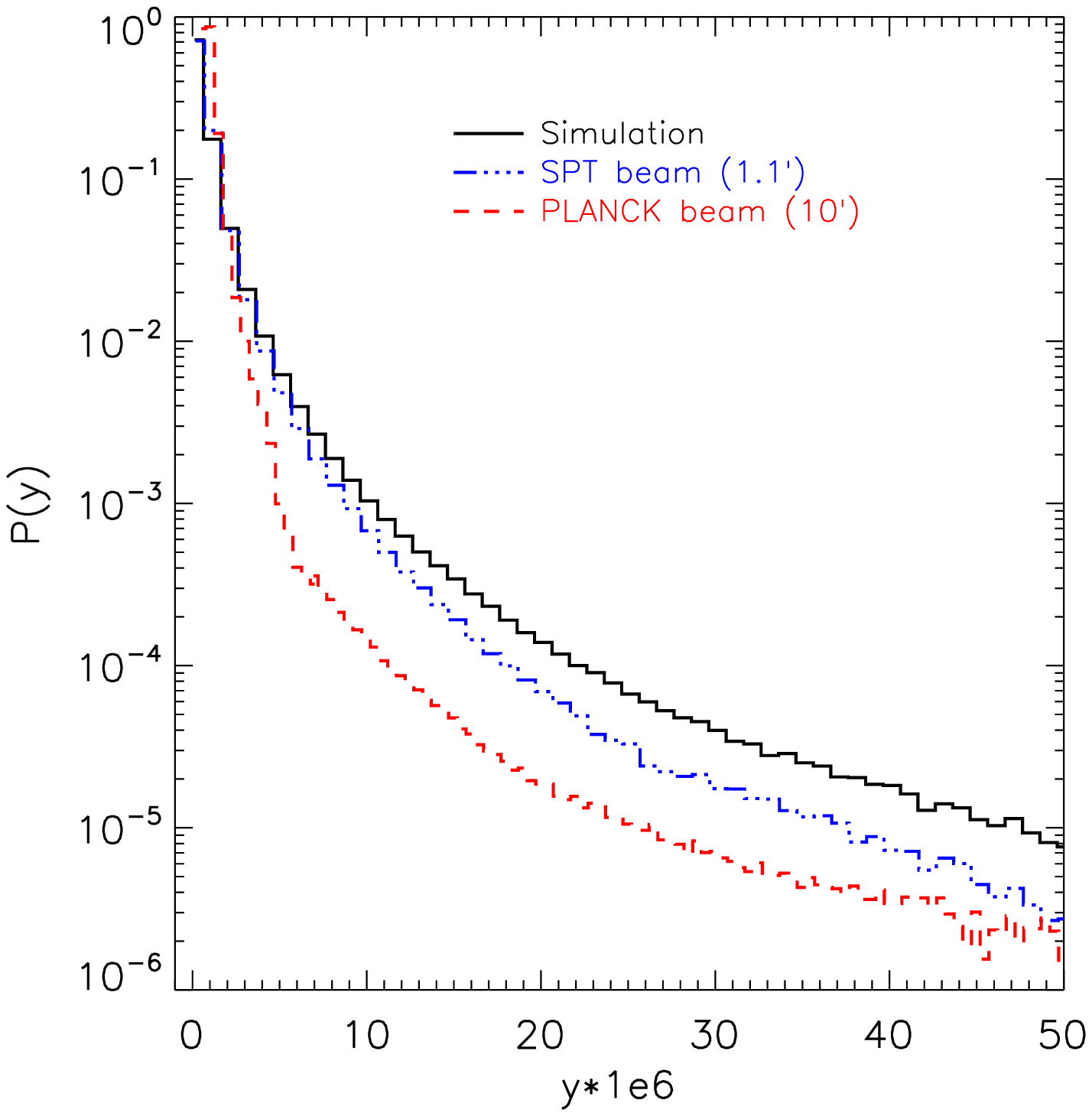}
\includegraphics[width=0.49\textwidth]{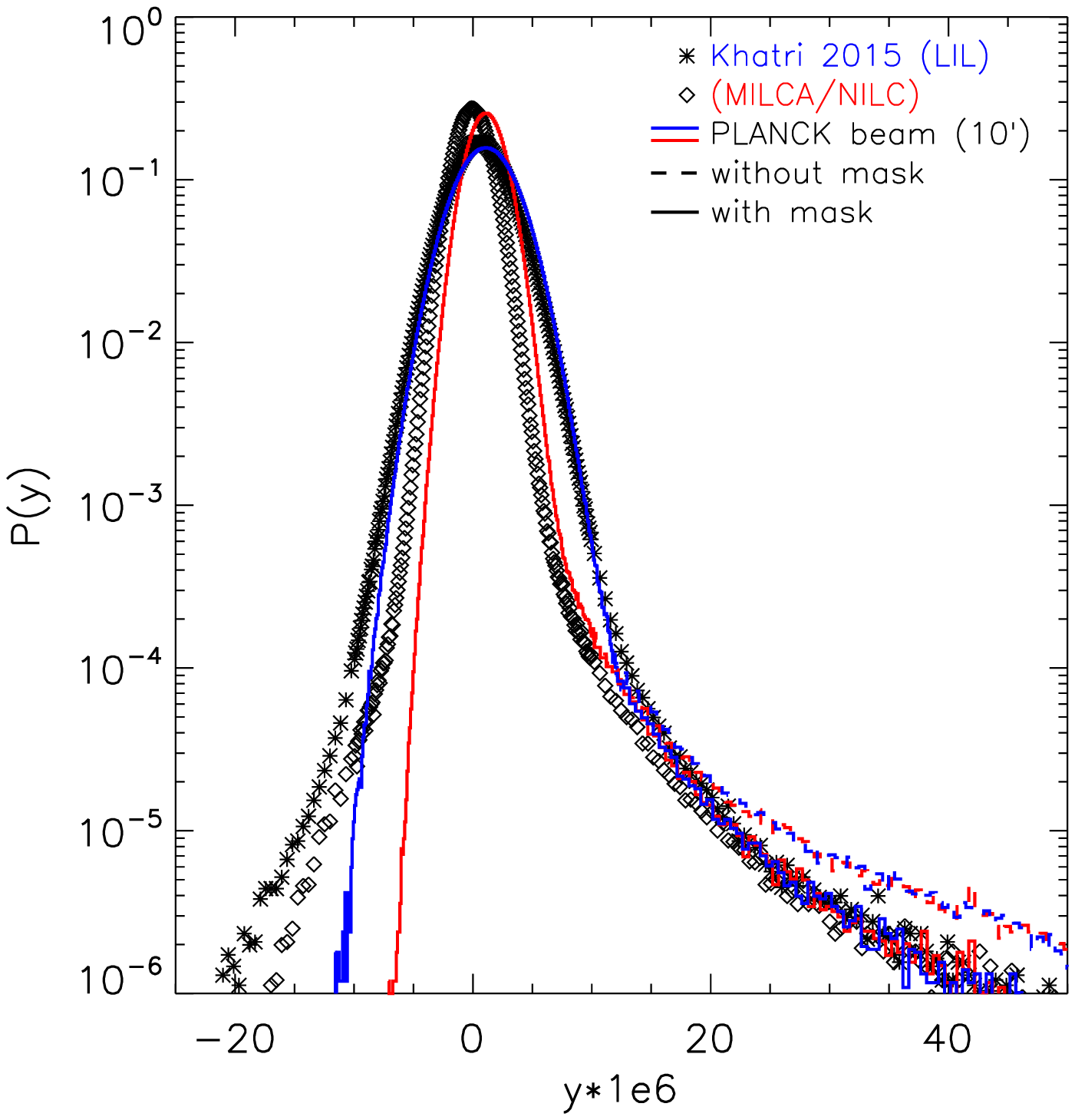}
\caption{
 PDFs of the Compton $Y$ parameter, normalised such that $\int
 d\hat{y}~P(\hat{y})=1$ with $\hat{y}=10^6y$.
(Left) The black line shows the PDF of the deep $8^\circ.8\times 8^\circ.8$ light-cone
 out to $z=5.2$ without any smoothing applied, while the blue dashed-triple-dotted line
 shows the PDF smoothed with 1.1 arcmin FWHM beam to simulate the signal
 that would be measured by SPT. The red dashed line shows the PDF including the local
 universe simulation, smoothed with 10 arcmin FWHM beam to simulate the signal
 that would be measured by Planck. (Right) Comparison to the Planck
 data. The black symbols show two estimates (``LIL'' and ``MILCA/NILC'')
 of PDFs from the Planck data
 \citep{2015arXiv150500781K,2015arXiv150500778K}. The ``LIL'' method
 yields larger noise in the Compton $Y$ map, and thus has a
 broader core in the middle, while both methods agree well in the tail.
 The red and blue lines show the PDFs from the simulation with Gaussian
 noise of $\sigma=1.5\times 10^{-6}$ (red) and $2.5\times 10^{-6}$
 (blue), which approximate the estimates of noise in LIL and MILCA/NILC maps,
 respectively. The excess at large $y$ values above the Planck data is due to the structures like Perseus-Pisces that are masked when the measurement is done. 
Applying the same mask with $f_{\rm sky}=0.51$ as used by
 \citet{2015arXiv150500781K}, we find that the PDFs
 from the simulation (solid lines) agree well with the Planck data at
 all values of $y$.
 }
\label{fig:pdf}
\end{figure*}

In the left panel of figure \ref{fig:coma}, we compare the radial
profiles of tSZ (the Compton $Y$ parameter given in
Eq.~\ref{eq:y_theta}), as well as the significant contribution
of kSZ at 150 GHz (as given in Eqs.~\ref{eq:w_theta} and \ref{eq:yeff}),
toward the Coma cluster in the simulation with the Planck tSZ
data for Coma \citep{2013A&A...554A.140P}. The blue line shows the Compton $Y$ toward the ``Coma
cluster'' in the local universe simulation out to a co-moving distance
of 110 Mpc, while the thick black
solid line shows the ``Coma cluster'' embedded in the deep light-cone out to
$z=5.2$. The latter has more signal in the outskirt of the cluster because
of the contributions from hot gas outside Coma.

We find an excellent agreement between the simulation and the Planck
data (shown as the symbols) in the inner part, $r\lesssim
100$~kpc. The full light-cone integration (thick black line) agrees
reasonably well with the Planck data up to a few Mpc. Given the
uncertainties in the constrained simulation, this level of agreement is
remarkable. Also the mean Compton $Y$ ($\bar Y=1.18\times 10^{-6}$) obtained from
the deep light-cone (shown by the horizontal dotted line) agrees with
the background subtracted from the Planck data (shown by the horizontal
dashed line) to within a factor of two. 

Finally, to get a feel for the magnitude of kSZ toward the ``Coma
cluster'', we show the sum of tSZ and kSZ at 150~GHz
(Eq.~\ref{eq:yeff}) in the magenta line. Both include the full
light-cone integration out to $z=5.2$. For this particular
realisation, we find a non-negligible (10 to 20 per cent) contribution
from kSZ.\footnote{However, we should not compare the magenta line
with the Planck data in the left panel of figure \ref{fig:coma}, as
the Planck data shown here are obtained by combining the
multi-frequency Planck data specifically to extract tSZ, and most
(if not all) of kSZ has been removed.} In particular, we find that
the local universe simulation predicts that the overall halo of Coma
is moving away from us at $\approx 400~{\rm km/s}$ with respect to the
CMB rest frame, yielding a negative kSZ signal up to Mpc radius. Being
a merging system, the core of Coma in our realisation moves with even
higher velocity (up to 800 km/s) as commonly seen in simulations
\citep{2010ApJ...717..908Z,2013MNRAS.432.1600D} near the center.

We also find a large, positive kSZ signal in the outskirt, which is due
to a  massive, gas-rich infalling sub-structure moving toward us. See
the right panel of figure~\ref{fig:coma} for a map of the mass-weighted
velocity field around Coma. 
While the overall halo velocity predicted for Coma should be accurate to
the extent of the precision of the constraints used by the local
universe simulation, details of sub-structures present in this
realization of the local universe are far outside the predictive power
of such a simulation. Therefore, the prediction for a positive kSZ in
the outskirt and the phase of movement of the core should be interpreted with caution.

\subsection{The Compton $Y$ Map}  \label{sec:ymap}
\subsubsection{Simulation results}

In figure~\ref{fig:pdf}, we compare the one-point PDF of the Compton
$Y$ map from the simulations (lines) with that observed by Planck
\citep[symbols;][]{2014A&A...571A..21P,2015arXiv150500781K,2015arXiv150500778K}.
The asymmetry of the left and right tails (skewness) of $P(y)$ is due to
galaxy clusters and groups along the lines of
sight, as predicted previously \citep{2001MNRAS.328..669Y,2003MNRAS.344.1155R}. In the absence of noise, the pixel
values in the simulations are always positive, whereas Planck's $Y$
values can have negative values due to noise (or unaccounted other sources).

In the right panel, we have smoothed the simulated Compton $Y$ map with
a Gaussian beam with 10 arcmin FWHM, and added Gaussian noise to the simulations with
 $\sigma=1.5\times 10^{-6}$ and $\sigma=2.5\times 10^{-6}$ to
 approximate the noise estimates from ``LIL'' and ``MILCA/NILC'' maps of
 the Compton $Y$, respectively  \citep[see figure 3
 in][]{2015arXiv150500781K,2015arXiv150500778K}. Since Planck cannot
 measure the absolute value of $Y$, there is ambiguity in the exact zero
 point of the PDF. While the simulation agrees well with the ``LIL''
 map, there seems a slight offset in the zero point with respect to the
 ``MILCA/NILC'' map.

As our maps have significantly higher spatial resolution than
Planck, we find a larger excess of high $Y$ values (black line in the
left panel) if not smoothed. This is driven by the
central parts of halos along the lines of sight. This can be best seen
in figure \ref{fig:cone}, especially in the zoom-in in the right panel.
However, once smoothed with the beam size of Planck's $Y$ map (10 arcmin
FWHM), the excess due to these structures unresolved by Planck 
is reduced (red dashed line in the left panel), and the simulation and the Planck data are in
much better agreement up to $Y\approx 2\times 10^{-5}$. The effect of
the beam smearing in the light-cone map can be seen visually in the left
most panel in the bottom panels of figure \ref{fig:map}. 
With the smoothing size as large as this, the excess PDF at larger
values of $Y$ above the Planck data is dominated by the nearby
structures that subtend large angles in the sky. We find that the excess
is dominated by the structures in the local universe simulation. 
Remarkably, we could identify the source of the excess PDF as the
structures such as Perseus-Pisces that are masked when the measurement
is done. Using the same mask that retains 51\% of the sky used by
\citet{2015arXiv150500781K},
we find an excellent agreement between the PDFs from the
simulation (solid lines in the right panel) and the Planck data at all values of $Y$. The ``core'' of the PDF
from the simulation at small values of $Y$ is dominated by the
structures in the deep light-cone beyond the local universe simulation.

We also convolve our map with the SPT beam (1.1 arcmin FWHM; magenta
line in the left panel). In this case, the excess at large Compton $Y$ values is only
mildly suppressed, demonstrating that SPT-like instruments would be able
to resolve almost all the contributions of structures
resolved by our simulation. See \citet{2014arXiv1411.8004H} for the
measurement and interpretation of the one-point PDF of the ACT data.

\begin{figure*}
\includegraphics[width=0.33\textwidth]{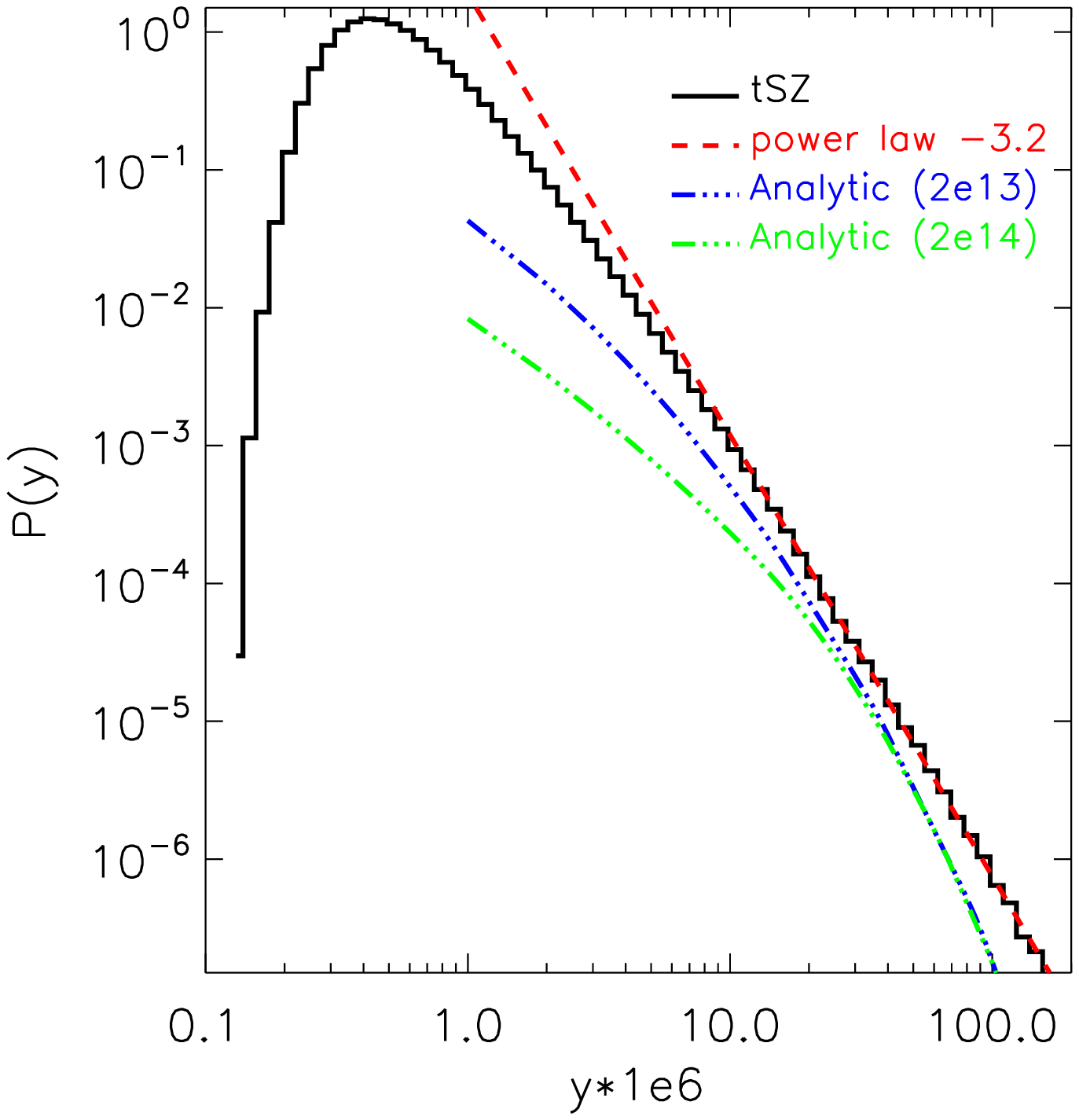}
\includegraphics[width=0.33\textwidth]{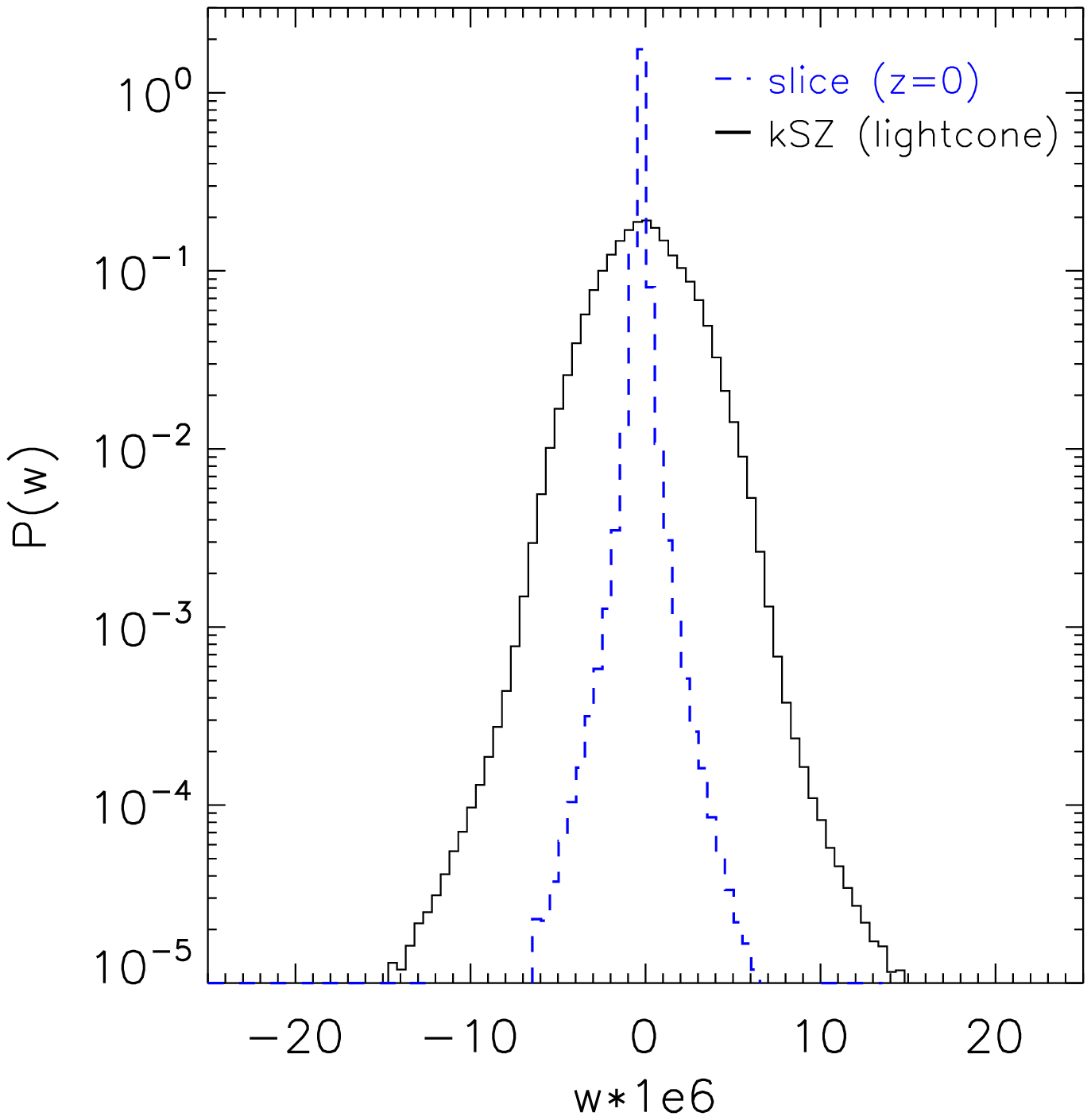}
\includegraphics[width=0.33\textwidth]{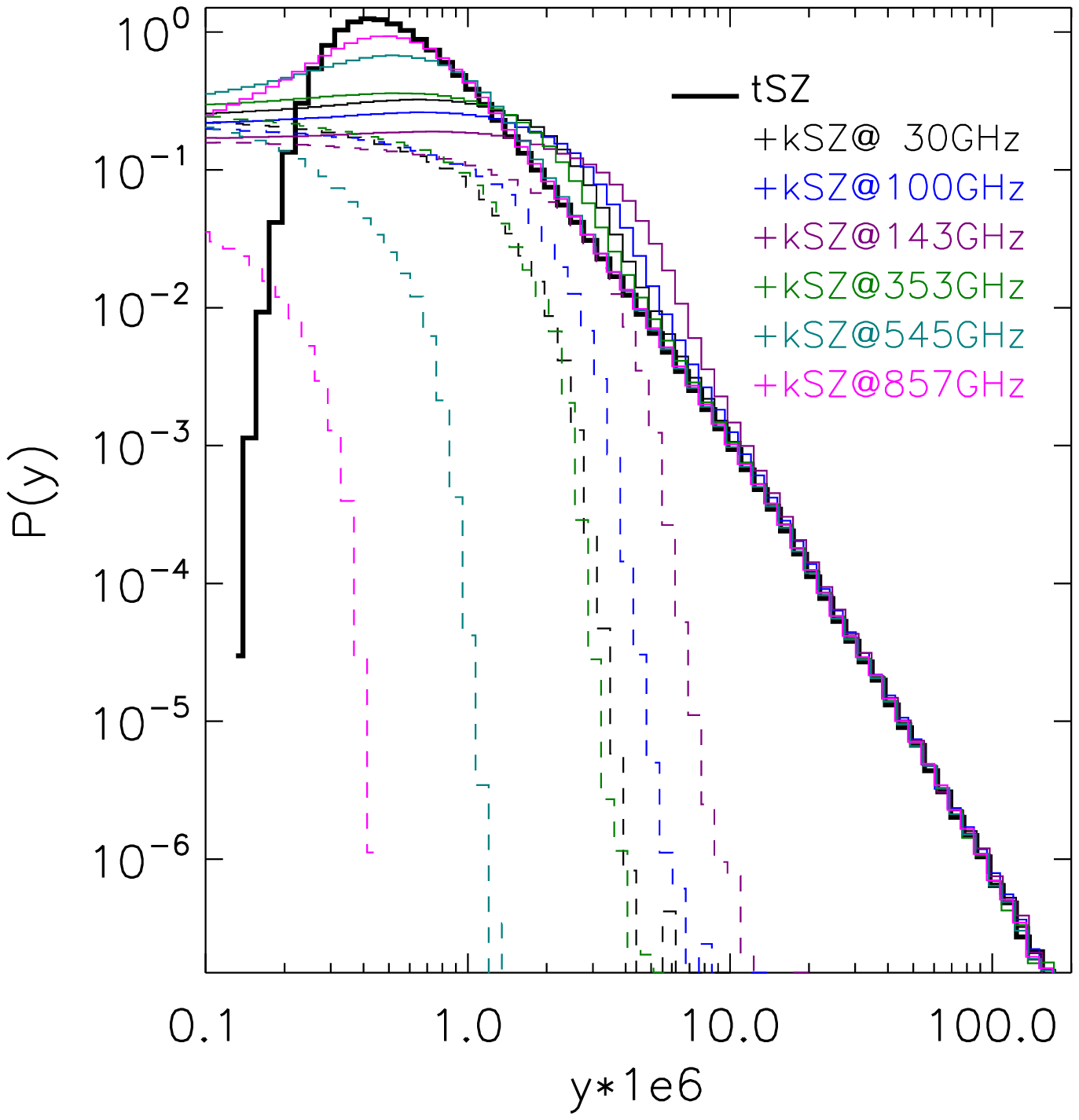}
\caption{PDFs of the SZ effects computed from the deep light-cone. (Left) PDF of the Compton $Y$ parameter, which is the same as the black line in figure~\ref{fig:pdf}, normalised such that $\int d\hat{y}~P(\hat{y})=1$ with $\hat{y}=10^6y$. The dashed line
 shows a power-law with a slope of $-3.2$. 
The analytical models with
 $M_{\rm min}=2\times 10^{13}$ and $2\times 10^{14}~M_\odot/h$
 are shown for comparison.
(Middle) PDF of kSZ,
 normalised such that $\int d\hat w~P(\hat w)=1$ with $\hat{w}=10^6w$.
The dashed line
 shows the PDF at $z=0$ \citep[see][for comparison]{2001MNRAS.328..669Y}, while the solid line shows the PDF from the
 light-cone. (Right) PDF
 of the sum of the two (Eq.~\ref{eq:yeff}) at various Planck
 frequencies, normalised such that $\int d\hat{Y}_{\rm eff}~P(\hat{Y}_{\rm eff})=1$ with $\hat{Y}_{\rm eff}=10^6Y_{\rm eff}$.
The solid lines show PDFs for $Y_{\rm eff}>0$ while the
 dashed lines show those for $Y_{\rm eff}<0$.}
\label{fig:cone_pdf}
\end{figure*}

Coming back to the unsmoothed PDF, we find that the tail of the PDF
follows a power law shape over at least two orders of magnitude in
Compton $Y$ values. The slope of this power law is approximately $-3.2$ as shown
by the dashed line in the left panel of figure~\ref{fig:cone_pdf}.
To quantify non-Gaussianity of the PDF, we calculate low-order
cumulants of $\hat y\equiv y\times 10^6$. We find $1.184$, $4.292$,
$24.27$, and $1514$ for the mean, variance, skewness, and kurtosis, respectively.

The middle panel shows the PDF of kSZ (solid line) as well as the
contribution of the $z=0$ slice. It shows a
non-Gaussian tail in agreement with the previous work
\citep{2001MNRAS.326..155D,2001MNRAS.328..669Y,2009ApJ...698.1795H}. 
The low-order cumulants of $\hat w\equiv w\times 10^{-6}$ are $0.17$,
$5.1$, $0.054$, and $0.71$ for the mean, variance, skewness, and kurtosis, respectively. The ensemble average of the mean should vanish;
a non-zero value from the simulation is due to cosmic variance.
The right panel shows
the PDF of the sum of tSZ and kSZ ($Y_\mathrm{eff}$ defined in
Eq.~\ref{eq:yeff}) at various Planck frequencies. The contribution to
the PDF of $Y_{\rm eff}$ of the kSZ signal is negligible for $Y_{\rm
eff}\gtrsim 10^{-5}$, while it significantly modifies the PDF at smaller
values of $Y_{\rm eff}$. As the PDF of kSZ is flatter at
small values of kSZ than that of the Compton $Y$ at small values of $Y$,
the PDF of the sum of the two shows an excess probability in
$10^{-6}\lesssim Y_{\rm eff}\lesssim 10^{-5}$ (see the solid lines
figure~\ref{fig:cone_pdf}). The dashed lines show the PDF for negative
values of $Y_{\rm eff}$. This unique shape of the PDF, which
changes as a function of frequencies in a predictable way, may be used
to detect the kSZ signals in the data.

\subsubsection{Analytical model of the PDF of Compton $Y$}
\label{sec:analytical_pdf}
The tail of the PDF of Compton $Y$ can be calculated analytically
following \citet{2014arXiv1411.8004H}. Specifically, we compute
\begin{align}
\nonumber
 P(Y)=\frac{\pi}{\Delta Y}&\int_0^5dz\frac{dV}{dz}\int_{M_{\rm min}}^{M_{\rm
  max}}d\ln M\frac{dn(M,z)}{d\ln
  M}\\&[\theta^2(M,z,Y)-\theta^2(M,z,Y+\Delta Y)]\,,
\label{eq:analytical_pdf}
\end{align}
where $\Delta Y=10^{-6}$, 
$M$ is the virial mass with $(M_{\rm min},\,M_{\rm max})=(5\times 10^{14}~M_\odot/h,\,5\times 10^{15}~M_\odot/h)$, $dV/dz$ is the differential comoving volume per
steradian, and $dn/d\ln M$ is the halo mass function given by 
\begin{equation}
 \frac{dn(M,z)}{d\ln M}=\frac{d\ln M_{200{\rm m}}}{d\ln
  M}\frac{dn(M_{200{\rm m}},z)}{d\ln M_{200{\rm m}}}\,,
\end{equation}
with $dn/d\ln M_{200{\rm m}}$ derived from simulations. Here, $M_{200{\rm m}}$ is the mass
enclosed within $r_{200{\rm m}}$, in which
the mean overdensity is $200$ times the mean mass density of the
universe. We convert $M$ to $M_{200{\rm m}}$ using an NFW profile
\citep{navarro/frenk/white:1996,navarro/frenk/white:1997} with
the concentration parameter of \citet{duffy/etal:2008} \citep[see,
 e.g., Eq.~14 of][]{komatsu/seljak:2001}. For the mass
range we consider, we find $d\ln M_{200{\rm m}}/d\ln M\approx 1$ to good
approximation.
For $dn/d\ln M_{200{\rm m}}$, we use the mass function from the Magneticum Pathfinder
simulation \citep[Eq.~1 and 3 of][with the parameters for ``$M_{200{\rm m}}$
Hydro'' given in their Table 2]{bocquet/etal:prep}.

To compute the linear r.m.s. mass density fluctuation necessary in the
fitting formulae of the mass function, 
we use the {\small CAMB} code \citep{lewis/challinor/lasenby:2000} to
generate the linear matter power spectrum with the cosmological
parameters of the Magneticum Pathfinder simulation.

The ``$\theta$'' in Eq.~\ref{eq:analytical_pdf} is the inverse function
of
\begin{equation}
 Y(\theta,M_{500},z)=\frac{\sigma_T}{m_e c^2}\int_{-6r_{500}}^{6r_{500}} dl~P_e(\sqrt{l^2+D_A^2\theta^2})\,,
\end{equation}
where $P_e(r)$ is the electron pressure profile and 
 $D_A$ is the proper angular diameter distance. That is to say, for
a given value of $Y$, $M_{500}$, and $z$, we find the corresponding value of
$\theta$ using this equation. $M_{500}$ is the mass enclosed inside $r_{500}$,
within which the mean
overdensity is 500 times the critical density of the universe. Again, we
convert $M$ to $M_{500}$ using an NFW profile with the concentration
parameter of \citet{duffy/etal:2008}. 

For $P_e$ we use the following
parametrized profile
\citep{nagai/kravtsov/vikhlinin:2007,arnaud/etal:2010}:
\begin{eqnarray}
\nonumber P_e(x) &=& 1.65~(h/0.7)^2~{\rm eV~cm^{-3}}\\ & \times&
E^{8/3}(z)\left[\frac{M_{500}}{3\times
    10^{14}(0.7/h)M_\odot}\right]^{2/3+\alpha_p} p(x),
\label{eq:pressure}
\end{eqnarray}
with $x\equiv r/r_{500}$, $\alpha_p=0.12$, $E(z)\equiv H(z)/H_0 =
\left[\Omega_m(1+z)^3+1-\Omega_m\right]^{1/2}$, and the function
$p(x)$ is defined by
\begin{equation}
p(x) \equiv \frac{6.41(0.7/h)^{3/2}}{(c_{500} x)^\gamma[1+(c_{500}x)^\alpha]^{(\beta-\gamma)/\alpha}},
\label{eq:gnfw}
\end{equation}
with $c_{500}=1.81$, $\alpha=1.33$, $\beta=4.13$, and $\gamma=0.31$
\citep{planck-int5:2013}.  However, this fitting function for the
pressure profile was derived for $M_{500}$ assuming hydrostatic
equilibrium, which is known to be biased low relative to the true
$M_{500}$ due to non-thermal pressure \citep[see, e.g.,][and
  references therein]{shi/komatsu:2014}. We thus rescale $M_{500}$  as
  $M_{500}\to M_{500}/1.2$ and $r_{500}$ as $r_{500}\to r_{500}/1.2^{1/3}$ to account for the hydrostatic mass bias. This
correction brings the pressure profiles in the Magneticum
  Pathfinder simulation into good agreement with the Planck data
\citep{planck-int5:2013}. 

Eq.~\ref{eq:analytical_pdf} is valid when halos included in the
calculation do not overlap in the sky. The overlapping fraction, $F_{\rm
clust}$, as defined by Eq.~10 of \citet{2014arXiv1411.8004H}, is 0.32
for $M_{\rm min}=2\times 10^{14}~M_\odot/h$. We show the result of the
analytical model in the left panel of figure~\ref{fig:cone_pdf}. The
shape of the PDF in the tail is reproduced reasonably well. We also show
the result for a smaller mass limit, $M_{\rm min}=2\times
10^{13}~M_\odot/h$, showing contributions from lower mass
halos. However, this case gives $F_{\rm clust}=5.5$, i.e., haloes
overlap significantly in the sky, and thus the calculation cannot be
trusted for lower values of $Y$. Nonetheless this calculation gives an
idea about the contributions from lower mass haloes. Since the PDF in
the extreme tail is sensitive to the core of the pressure profiles, a
difference in details of the pressure profile in the simulation and that
used in the analytical model shows up there: the simulation gives more
pressure in the cores of galaxy clusters than that of Eq.~\ref{eq:pressure}.

\subsection{The angular power spectra of tSZ and kSZ} \label{sec:angpower}
\subsubsection{Simulation results}

To calculate the angular power spectrum of tSZ, we separately
use the full-sky map obtained from the local universe simulation in
$z<0.027$ and the light-cones over
$8^\circ.8\times8^\circ.8$ in $0.027<z<5.2$.
As these simulations are performed with
different cosmological parameters, we rescale the amplitudes of
the tSZ power spectra by $\sigma_8^8\Omega_m^3$ to Planck 2015's best fitting
CMB cosmological values, $\Omega_m=0.308$ and
$\sigma_8=0.8149$ \citep[``TT+lowP+lensing''
of][]{2015arXiv150201589P}. We have checked that this scaling agrees
with the scaling predicted by the analytical calculation presented in section~\ref{sec:analytical}

Figure~\ref{fig:power} shows the tSZ power spectrum measured by Planck
at 150~GHz \citep[red symbols with error bars;][]{2015arXiv150201596P}. 
Two dashed lines show the tSZ power spectrum from the local
universe at low multipoles and that from the light-cone at high
multipoles, while the black solid line shows the sum of the two.
This division in multipoles is consistent with the previous work showing
that the nearby structures dominate at low multipoles simply because
they appear larger in the sky \citep{refregier/etal:2000,komatsu/seljak:2002,dolag2005,2005MNRAS.361..753H}.

The black solid line agrees with the Planck data well at all multipoles
measured by Planck, i.e., $l\lesssim 1000$. Our conclusion that the predicted tSZ power spectrum
with Planck 2015's TT+lowP+lensing parameters agrees with the
measured power spectrum is consistent with the finding of the Planck
team \citep{2015arXiv150201596P}. 
\citet{2014MNRAS.440.3645M} show that
the Planck 2013 parameters with $\Omega_m=0.3175$ and $\sigma_8=0.834$
\citep{2014AA...571A..16P}
over-predict the tSZ power spectrum, and show that another set of
parameters with $\Omega_m=0.302$ and $\sigma_8=0.817$
\citep{2015PhRvD..91b3518S} gives the tSZ power spectrum in agreement
with the measurement. The latter set is close to the Planck 2015
parameters with lensing that we use in this paper; thus, our conclusion
is consistent with their results. Using the $\Omega_m^3\sigma_8^8$
scaling, for example, the former set gives 32\% larger power than the
2015 parameters, while the latter gives 4\% smaller power.

However, at $l\approx 3000$, our prediction is significantly higher than
the measurements reported by SPT and Atacama Cosmology Telescope (ACT)
collaborations \citep{2012ApJ...755...70R,2013JCAP...10..060S}. This
finding is not new: the previous calculations of the tSZ power spectrum also
over-predict the power at $l\approx 3000$ compared to the SPT and ACT
data, although the degree of overestimation varies depending on the
details of baryonic physics implemented in the models
\citep[e.g.,][]{2014MNRAS.440.3645M,2015ApJ...799..177G,2014arXiv1412.6023R}. 

Whether this discrepancy at $l\approx 3000$
poses a serious challenge to theory is unclear, given that the SPT and ACT do
not have as many frequency channels as Planck. Distinguishing the
primary CMB, tSZ, and the other extra-galactic sources by their
frequency dependence is more challenging for SPT and ACT. (On the other hand, Planck
does not have angular resolution to resolve the power at $l\approx 3000$.)
The tSZ signal is a sub-dominant contribution to the power spectrum at
$l\approx 3000$ compared to the extra-galactic
sources. The
magenta symbols with error bars in figure~\ref{fig:power} show the SPT
power spectrum with the best-fitting primary CMB power spectrum
subtracted. The difference between the magenta symbols and a magenta
vertical line at $l=3000$ indicates the amount of the extra-galactic
power that needs to be subtracted. At least, our tSZ power spectrum does
not overshoot the magenta data points that provide a firm upper
limit on the tSZ power at these angular scales.

Next, we show the kSZ power spectrum from the light-cones by 
the blue solid line in figure~\ref{fig:power}. Our result at high multipoles
($l\gtrsim 1000$) agrees with that of \citet{2012ApJ...756...15S}. The upturn
at lower multipoles can be understood by the contribution from the
longitudinal velocity contributions that did not fully cancel by the
line-of-sight integration \citep{2009MNRAS.398..790H}. At 150~GHz, the
kSZ amplitude becomes comparable to tSZ at $l\approx 300$, and becomes
even dominant at lower multipoles. (Note that the Planck data shown in
this figure remove most of the kSZ signals by construction, and thus
should not be compared with the blue line.)
 However, as the kSZ signal is
dominated by the largest modes present in the simulation (see
figure~\ref{fig:map}), even larger cosmological volumes will be needed
to obtain a fully converged kSZ results from such light-cones.

\subsubsection{Analytical model of the tSZ power spectrum}
\label{sec:analytical}
To check accuracy of scaling the tSZ power spectrum to other
cosmological parameters, we compute the tSZ power spectrum using an
analytical model. Ignoring a small contribution from the correlation
between two distinct dark matter halos \citep{komatsu/kitayama:1999}, we
model the SZ power spectrum as \citep{komatsu/seljak:2002}
\begin{equation}
 C_l=g^2(x)\int_0^5dz\frac{dV}{dz}\int_{M_{\rm min}}^{M_{\rm
  max}}d\ln M\frac{dn(M,z)}{d\ln M}|\tilde{y}_l(M,z)|^2\,,
\label{eq:clsz}
\end{equation}
where $(M_{\rm min},\,M_{\rm max})=(5\times 10^{11}~M_\odot/h,\,5\times
10^{15}~M_\odot/h)$. 
The 2D Fourier transform of the Compton $Y$ parameter, $\tilde{y}_l$, is
given by
\begin{equation}
 \tilde{y}_l=\frac{4\pi
  r_{500}}{l_{500}^2}\frac{\sigma_T}{m_ec^2}\int_0^6 dx x^2
  P_e(x)\frac{\sin(lx/l_{500})}{lx/l_{500}}\,,
\label{eq:yl}
\end{equation}
where $x\equiv r/r_{500}$ and $l_{500}\equiv D_A/r_{500}$. The electron pressure
profile, $P_e(x)$, is given by Eq.~\ref{eq:pressure}.

\begin{figure}
\includegraphics[width=0.49\textwidth]{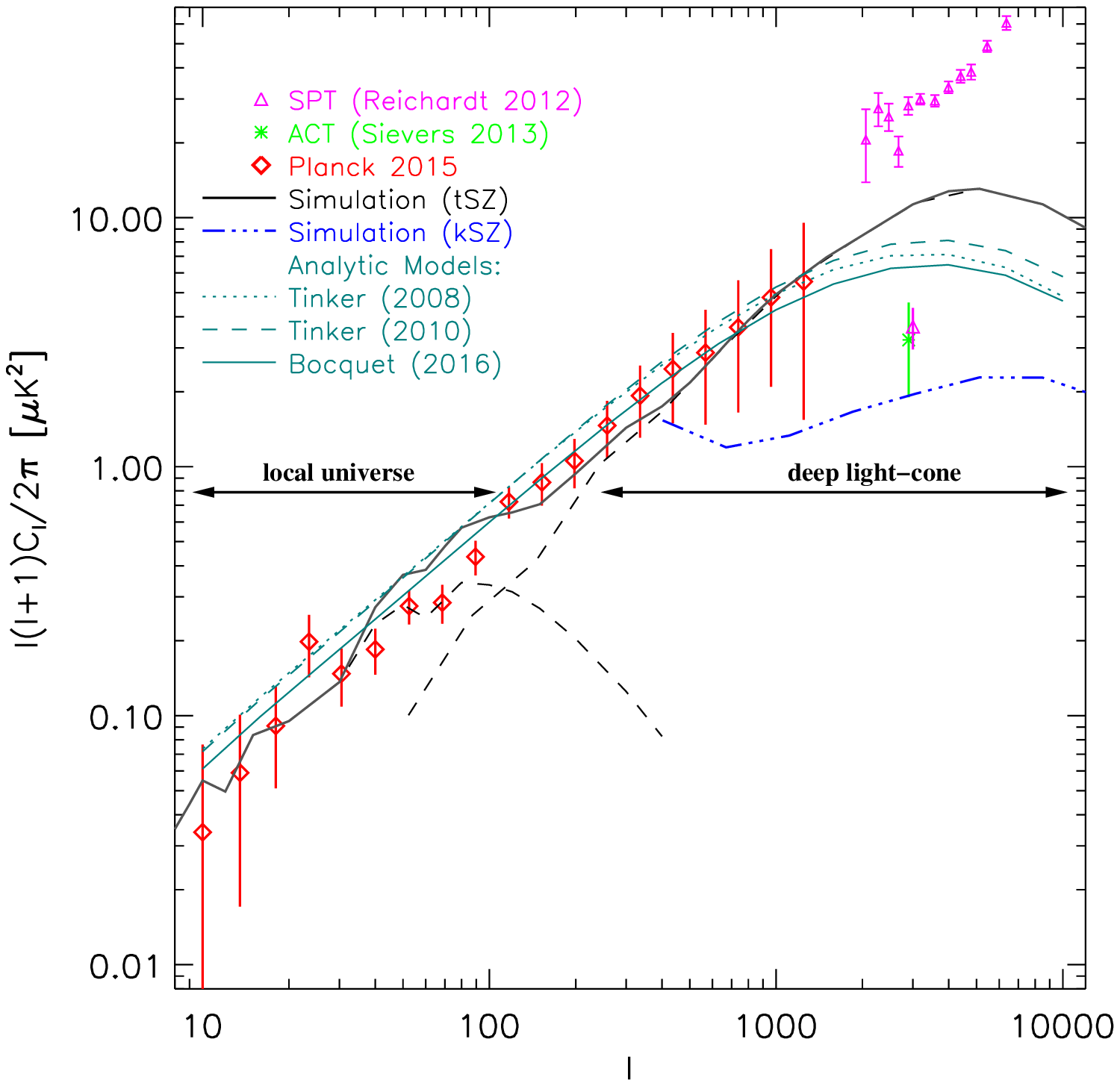}
\caption{Power spectra of temperature anisotropies due to tSZ and kSZ at
 150~GHz in units of $\mu{\rm K}^2$. The red symbols with error bars show the estimation of the tSZ power spectrum
 from the Planck data \citep{2015arXiv150201596P}, while the magenta
 symbols with error bars show the power spectrum of the SPT data
 \citep{2012ApJ...755...70R} with  the best-fitting primary CMB power
 spectrum subtracted. The green and magenta vertical lines show the
 ranges of the estimated tSZ power spectra at $l=3000$ by ACT
 \citep{2013JCAP...10..060S} and SPT, respectively. 
The black dashed lines show the tSZ power spectrum of the full-sky, local
 universe simulation ($z<0.027$) at low multipoles, and that of the deep,
 $8^\circ.8\times8^\circ.8$ light-cone from the Magneticum Pathfinder
 simulations ($0.027<z<5.2$) at high multipoles. The black solid line
 shows the sum of the two.
Both power spectra are scaled to Planck 2015's ``TT+lowP+lensing''
 cosmological parameters of $\Omega_m=0.308$ and $\sigma_8=0.8149$. The
 light blue lines show the analytical predictions based on three
 different mass functions (see section~\ref{sec:analytical}). The tripple dot-dotted
 blue line shows the kSZ power spectrum of the deep light-cone.}
\label{fig:power}
\end{figure}

For $dn/d\ln M_{200{\rm m}}$, we use three different sets of fit
parameters obtained from numerical simulations in the literature. First,
we use the mass function from the Magneticum Pathfinder
simulation \citep[Eq.~1 and 3 of][with the parameters for ``$M_{200{\rm m}}$
Hydro'' given in their Table 2. Using ``$M_{200{\rm m}}$
DMonly'' gives a similar result: the difference in the power spectrum is
less than four percent at all multipoles]{bocquet/etal:prep}.
This mass function
should give the result that is most consistent with the black solid line
shown in figure~\ref{fig:power}. The other mass function fits are
taken from  \citet{2008ApJ...688..709T} and
\citet{2010ApJ...724..878T}. 

To compute the linear r.m.s. mass density fluctuation necessary in the
fitting formulae of the mass function, 
we use the {\small CAMB} code \citep{lewis/challinor/lasenby:2000} to
generate the linear matter power spectrum with the Planck 2015
``TT+lowP+lensing'' parameters: 
$\Omega_bh^2=0.02226$,
$\Omega_ch^2=0.1186$, $\Omega_\nu h^2=0.00064$, $h=0.6781$, $\Delta_{\cal
R}^2(0.05~{\rm Mpc}^{-1})=2.139\times 10^{-9}$, and $n_s=0.9677$ \citep{2015arXiv150201589P}.

The light blue solid line shows in Fig. \ref{fig:power} the analytical model calculation with
the mass function of \citet{bocquet/etal:prep}, which is in good
agreement with the results obtained directly from the simulation at
$l\lesssim 1000$, while it under-predicts the power at
higher multipoles. This can be understood partially as due to inhomogeneity in
the distribution of pressure within halos, as the analytical model
assumes a smooth distribution of pressure. \citet{2012ApJ...758...75B}
show that inhomogeneity increases the power at these high multipoles by
20\%. The remaining difference may be due to the pressure profile of
low-mass halos (less than $10^{14}~M_\odot$) in the Magneticum Simulation
being slightly different from the Planck pressure profile given in
Eq.~\ref{eq:pressure}, as well as to the normalisation of the mass function for
low-mass halos (see the next paragraph). In any case, the overall
agreement between the simulation and the analytical model is satisfactory.

The analytical models with the other mass functions yield similar, but
different, results. The only difference between the mass functions of
\citet{2008ApJ...688..709T} and \citet{2010ApJ...724..878T} is that the
latter forces the normalisation of the mass function by $\int_0^\infty
dM_{200{\rm m}}~ dn/d\ln M_{200{\rm m}}=\bar\rho$, where $\bar\rho$ is the mean mass density of the
universe. This normalisation mainly changes the abundance of low-mass
halos which are not well resolved by their N-body simulations. As a
result, the latter mass function (shown as the dashed light blue line in
figure~\ref{fig:power}) gives larger power at high multipoles where the
contributions from low-mass halos dominate. 

At $l\lesssim 100$, the tSZ power spectra with both Tinker et al. mass
functions are slightly larger than that with the mass function of
\cite{bocquet/etal:prep} (20 and 15 per cent larger at $l=10$ and 100,
respectively).
Indeed, the fitting formula for the mass
function of \cite{bocquet/etal:prep} gives similarly smaller $dn/d\ln M_{200{\rm m}}$
than the fits of Tinker et al. mass functions at the relevant mass scales,
i.e., $M_{200{\rm m}}\gtrsim 10^{15}~M_\odot$. \cite{bocquet/etal:prep} explain this by the
way the fits are performed; namely, the actual data of the N-body
simulations are similar between Bocquet et al. and Tinker et al., but the fitting
procedures give slightly different results. Bocquet et al. use
the Bayesian likelihood approach taking into account properly the
Poisson nature of the mass function measured from the simulation, whereas Tinker et al. use the $\chi^2$
statistics. The former results do not depend on the bin size of the mass
in which the fits to $dn/d\ln M_{200{\rm m}}$ are performed, whereas the latter
results do. Our results highlight the importance of better understanding
the high-mass end of the mass function for the study of the tSZ power spectrum.

\section{Mean Compton $Y$} \label{sec:evol}
\subsection{Simulation results}
\begin{figure}
\includegraphics[width=0.49\textwidth]{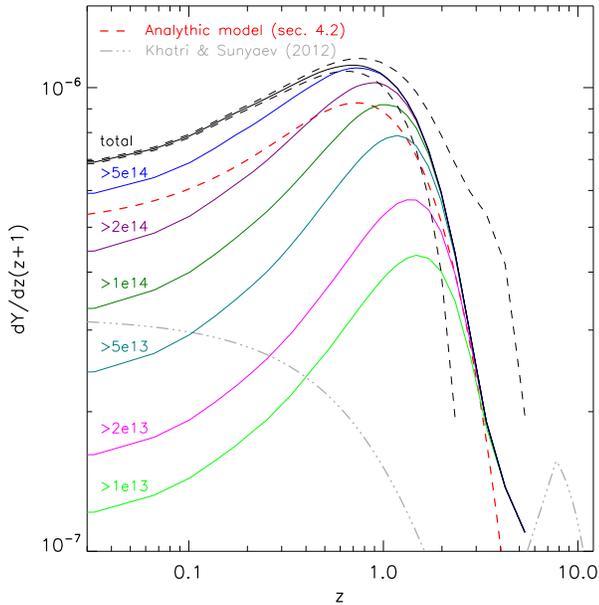}
\caption{How the mean Compton $Y$ builds up over time, d$\bar
 Y$/dln$(1+z)$. All redshifts below $z\approx 1.5$ contribute to the
 total signal (black solid line) almost equally, to within 30\%. The
 coloured lines show d$\bar Y$/dln$(1+z)$ with high-mass halos
 (whose virial masses are indicated by the numbers in units of $M_\odot/h$) removed from the
 simulation. Half of the signal at $z=0$ comes from clusters with
 $M>10^{14}~M_\odot/h$, whereas at $z\approx 1$ the bulk of the signal
 comes from lower-mass halos. The dashed lines show the scatter due to kSZ at
 150GHz. The bulk of kSZ comes from high redshift and non-collapsed regions.
 The gray solid line show approximate estimates of tSZ from the epoch of
 reionisation as well as from the intergalactic medium (IGM) at lower
 redshift taken from \citet{2012JCAP...09..016K}. The gray dashed line
 shows the analytical model for the contributions from halos.}
\label{fig:mean}
\end{figure}

The deep light-cone yields the mean Compton $Y$ of $\bar Y=1.18\times
10^{-6}$ (shown as the dotted horizontal line in figure~\ref{fig:coma}) for the cosmological parameters used in the Magneticum
Pathfinder simulation: $\sigma_8 = 0.809$ and $\Omega_m = 0.272$.
We now study how $\bar Y$ builds
up over cosmic time and how much objects of different masses contribute
to $\bar Y$. To this end, we proceed similar to the construction of the
different slabs contributing to the light-cone. However, to avoid that
the volume (and therefore the statistics) decreases with decreasing
redshift like the volume swept by a light-cone of a given
area in the sky, we produce maps of the full simulation at each
redshift.
This allows us to compute $\mathrm{d}\bar Y/\mathrm{d}z$ at every snap
shot with the same precision given by the comoving box size of our
simulation, which is slightly more than 2~Gpc$^3$. 

Figure~\ref{fig:mean} shows d$\bar Y$/dln$(1+z)$ as a function of $z$,
i.e., the time evolution of the contribution per logarithmic
redshift interval to the overall signal. We find that all redshifts
below $z\approx 1.5$ contribute almost equally to within 30\%. To study
which masses contribute, we also show d$\bar Y$/dln$(1+z)$ with
high-mass halos above a certain mass threshold removed from the
simulation. At $z=0$, half of the signal comes from
$M>10^{14}~M_\odot/h$, whereas at $z\approx 1$ smaller halos dominate.
We find that, even at $z=0$, there is more than 10\% of d$\bar
Y$/dln$(1+z)$ coming from outside of resolved objects, e.g., the diffuse
baryon component. This fraction increases at higher redshifts and reaches
almost 30\% at $z\approx1$.

We also show an order-of-magnitude comparison (gray solid line) by
\citet{2012JCAP...09..016K} of the contribution from the epoch of
reionisation and from low redshift WHIM. The first one is based on the
reionisation optical depth inferred from CMB observations by WMAP
whereas the later one is based on the simulations of \citet{1999ApJ...514....1C} and assumes that the
temperature of IGM is $10^4$~K at $z>3$ and $10^6/(1+z)^{3.3}$~K at $z<3$
and does not include the contribution of massive clusters of galaxies.
While these rough estimates cannot be compared quantitatively with the
simulation results, it demonstrates that these contributions are much smaller
than those from hot gas in galaxy clusters in the simulation.

The dashed gray line shows the analytical
model of the contributions of halos computed with the mass function of 
\citet{2010ApJ...724..878T} and the Planck pressure profile with the
mass bias of 1.2. The analytical model
does a reasonable job describing the simulation result, although it is
systematically lower than the simulation by 20\% at $z\lesssim 1$.
See section 4.2 for more detailed discussion on the analytical model.

The dashed black lines show the scatter due to kSZ at 150~GHz. The kSZ is 
most prominent at high redshift and starts to contribute even earlier
than tSZ does.

\begin{figure}
\includegraphics[width=0.49\textwidth]{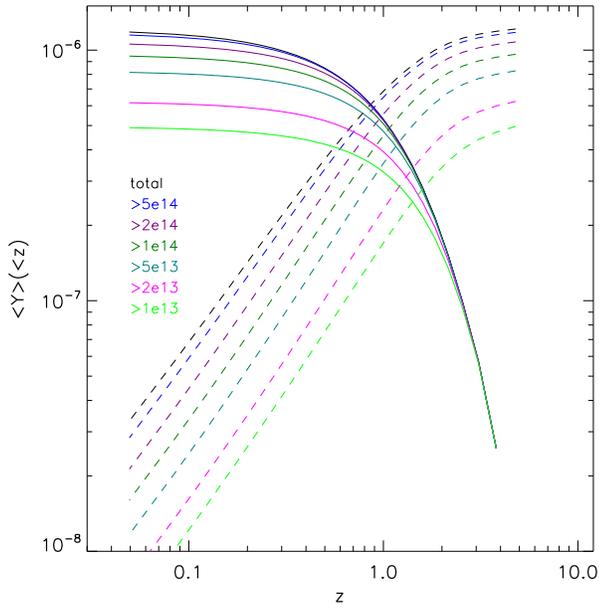}
\caption{Same as figure~\ref{fig:mean}, but for the cumulative signals. The solid lines show $\bar
Y(>z)=\int_z^{5.2}dz'~{\rm d}\bar Y/{\rm d}z'$, while the dashed lines
show $\bar Y(<z)=\int_0^{z}dz'~{\rm d}\bar Y/{\rm d}z'$.}
\label{fig:mean_buidup}
\end{figure}

In figure \ref{fig:mean_buidup}, we show the evolution of the cumulative
mean Compton $Y$. The solid lines show $\bar
Y(>z)=\int_z^{5.2}dz'~{\rm d}\bar Y/{\rm d}z$, while the dashed lines
show $\bar Y(<z)=\int_0^{z}dz'~{\rm d}\bar Y/{\rm d}z$. The latter
clearly shows that the mean Compton $Y$ receives significant
contributions up to $z\approx 2$. We find that cluster-size halos with
$M>10^{14}~M_\odot/h$ contribute only 20\% of the total signal, and
nearly half of the signal, $\bar Y=5\times 10^{-7}$, comes
from $M<10^{13}~M_\odot/h$ \citep[which can be detected by
stacking;][]{2013A&A...557A..52P,2014MNRAS.445..460G,2015ApJ...808..151G} and
the diffuse baryons outside halos.
  
\subsection{Analytical model}

The analytical model for the mean Compton $Y$ is given analogously to
Eq.~\ref{eq:clsz} \citep{barbosa/etal:1996}
\begin{equation}
 \bar Y=\int_0^5dz\frac{dV}{dz}\int_{M_{\rm min}}^{M_{\rm
  max}}d\ln M\frac{dn(M,z)}{d\ln M}\tilde{y}_0(M,z)\,,
\label{eq:meany}
\end{equation}
where $\tilde{y}_0$ is Eq.~\ref{eq:yl} with $l=0$. However, as the mean
Compton $Y$ receives significant contributions from lower mass halos
compared to the power spectrum, we use $M_{\rm
min}=5\times 10^{10}~M_\odot/h$. As our goal in this section is to confirm the
result of the previous section, we use the cosmological parameters of
the Magneticum Pathfinder simulation. All the other details of the
calculation are the same as in section~\ref{sec:analytical}.

We find $\bar Y=(0.78,0.83,0.99)\times 10^{-6}$ for the mass functions
of \citet{bocquet/etal:prep}, \citet{2008ApJ...688..709T}, and
\citet{2010ApJ...724..878T}, respectively. The mean Compton $Y$
receives significant contributions from low-mass halos for which these
mass functions differ significantly. In particular, the former two
fitting functions do not satisfy the normalisation constraint on the
mass function, $\int_0^\infty
dM_{200{\rm m}}~ dn/d\ln M_{200{\rm m}}=\bar\rho$.
A reasonable agreement between the results from the simulation
and the analytical model with the mass function of
\citet{2010ApJ...724..878T}, which does satisfy the normalisation
constraint, is encouraging; however, more study on a low-mass end of the
mass function is necessary. The remaining difference of order 20\%
relative to the simulation is due to the Planck pressure profile with
the mass bias of 1.2 being slightly lower than the pressure profiles in
the simulation in lower mass halos ($M\lesssim 10^{14}~M_\odot$). The
same trend can be seen in the tSZ power spectrum at $l\gtrsim 3000$
shown in figure~\ref{fig:power}. For example, we find $\bar Y=(1.07,1.16)\times
10^{-6}$ with the mass biases of 1.15 and 1.1, respectively. However,
the mass bias of 1.1 would yield too large a tSZ power spectrum at low
multipoles, $l\lesssim 1000$, to agree with the simulation.

\begin{figure}
\includegraphics[width=0.49\textwidth]{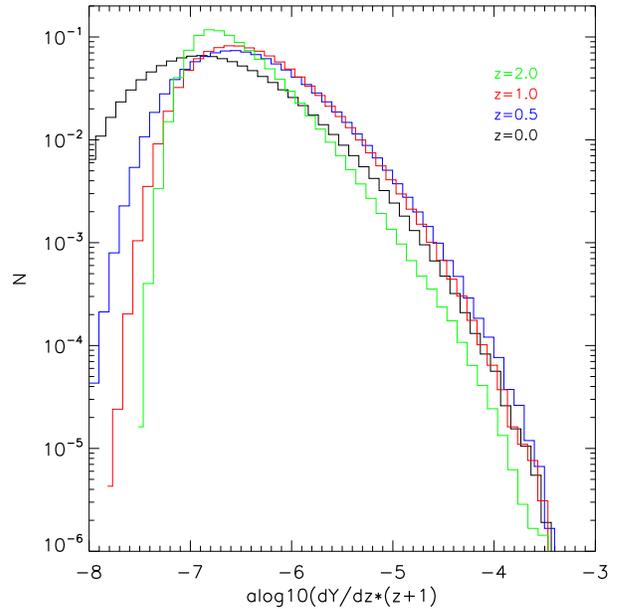}
\caption{PDFs of d$Y$/dln$(1+z)$ at $z=0$ (black), $0.5$ (blue),
 $1$ (red), and $2$ (green).}
\label{fig:distr_tot_evol}
\end{figure}

Using the Planck 2015 parameters, the mass function of
\citet{2010ApJ...724..878T} and the mass bias of 1.2, we find $\bar
Y=1.32\times 10^{-6}$. Using this to scale the simulation result, we
find $\bar Y=1.57\times 10^{-6}$.

Recently, \citet{2015arXiv150701583H} use the analytical model with the
mass function of \citet{2010ApJ...724..878T} and the pressure profile of
\citet{2012ApJ...758...75B} to obtain $\bar Y=1.58\times 10^{-6}$ for
the WMAP9 parameters with $\sigma_8=0.817$ and $\Omega_m=0.282$
\citep{2013ApJS..208...19H}.
A few minor details on the calculation are
different: while they integrate the pressure profile out to
2.5 times the virial radius, we integrate out to $6r_{500}$; 
their redshift integration is from $z=0.005$ to 6; and the lower boundary of their mass integration is $M_{\rm min}=5\times 10^5~M_\odot/h$. 
Changing the
upper integration boundary from $x_{\rm
max}=6$ to $2.5r_{\rm vir}/r_{500}$ in Eq.~\ref{eq:yl} and adjusting the integration boundaries in Eq.~\ref{eq:meany}, we find $\bar
Y=1.11\times 10^{-6}$ for the Planck pressure profile with
the mass bias of 1.2 and the same WMAP9 parameters. 
This value is significantly lower
than their value. We find that this is due to the difference in the
pressure profiles; while the pressure profiles of the Magneticum
Pathfinder simulation, the Planck pressure profile with the mass bias of
1.2, and the profile of \citet{2012ApJ...758...75B} agree in the
high-mass end, $M_{500}\gtrsim 2\times 10^{14}~M_\odot/h$
\citep{2013A&A...550A.131P}, the latter profile gives significantly
larger pressure than the Planck profile in lower masses that dominate in
$\bar Y$.

These discrepancies need to be understood before we obtain an accurate
estimate of the expected level of $\bar Y$. Nevertheless, the original conclusion
from the previous generation of cosmological hydrodynamical simulations \citep{refregier/etal:2000,2000MNRAS.317...37D,seljak/etal:2001,2004MNRAS.355..451Z} seems robust: the expected $\bar
Y$ from the large-scale structure is of order $10^{-6}$ and is only one
order of magnitude lower than the FIRAS bound.
It is also encouraging
that none of these estimates exceed the new Planck bound on the
fluctuating part of the mean $Y$ parameter, $\bar Y< 2.2\times 10^{-6}$
\citep{2015arXiv150500781K}.

\subsection{Buildup of the Compton $Y$ PDF} \label{sec:distr}

Finally, we study contributions from different redshifts
to the build up of the PDF of the Compton $Y$ signal.
Figure~\ref{fig:distr_tot_evol} shows the PDF of d$Y$/dln$(1+z)$ at
four different redshifts. Although these PDFs look at first glance similar
to the one of the full light-cone (as shown in
figure \ref{fig:cone_pdf}), the tail does not follow a simple power
law. In general, with decreasing redshift, the PDF becomes broader and
less sharply peaked, especially when compared to the PDF at $z=2$. 

\begin{figure*}
\includegraphics[width=0.99\textwidth]{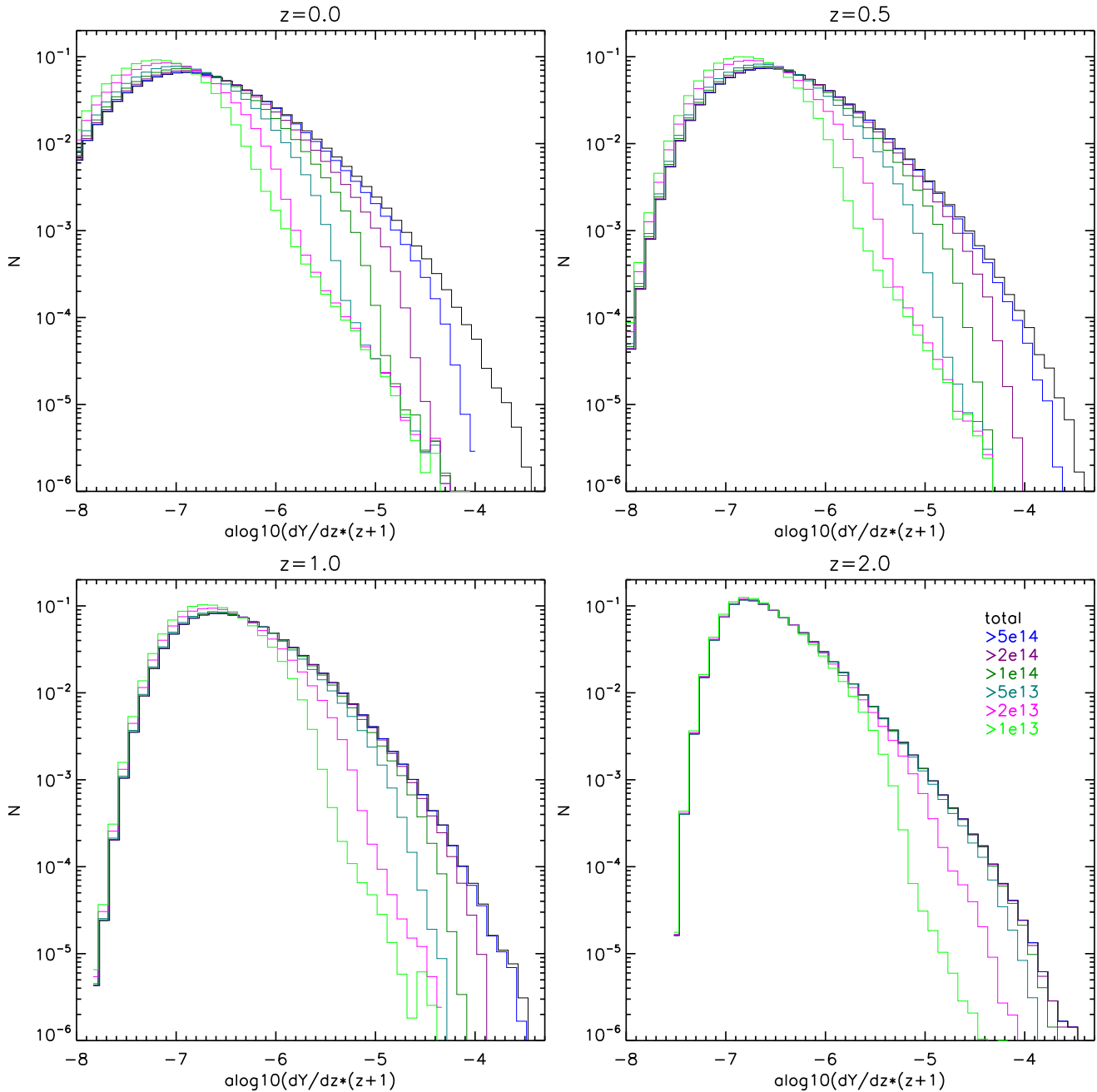}
\caption{PDFs of d$Y$/dln$(1+z)$ at $z=0$ (top left), $0.5$ (top right),
 $1$ (bottom left), and $2$ (bottom right). In each panel,
 we show PDFs  with high-mass halos above a certain mass threshold
 removed from the  simulation. The colours indicate the same mass thresholds
 as shown in figure \ref{fig:mean}.
}
\label{fig:distr}
\end{figure*}

In figure \ref{fig:distr}, we show PDFs with high-mass halos above a
 certain mass threshold removed from the simulation. At any given time,
 the tail is dominated by the most massive halos of that time.


\section{Conclusions}
The Magneticum Pathfinder simulations, state-of-the-art cosmological and
hydrodynamical simulations, follow in detail the thermal and chemical
evolution of the ICM as well as the evolution of super-massive black
holes and their associated feedback processes. These simulations
reproduce the average ICM pressure profiles measured by Planck
\citep{2013A&A...550A.131P} and SPT \citep{2014ApJ...794...67M}. At the
same time, the stellar mass functions of galaxies and the luminosity
functions of the AGN population agree well with observations
\citep{2014MNRAS.442.2304H}. The improved numerical methods and
increased computing power available today have enabled these simulations
to follow a large enough cosmological volume to construct realistic light-cone
maps, which can be compared with observations in detail. 

In this paper, we have computed the tSZ and kSZ effects toward the
counterpart of the Coma cluster in the local universe simulation, and
statistics of the SZ effects from the light-cone maps, including the
one-point PDF and power spectrum of  tSZ and kSZ, and the
mean Compton $Y$ parameter. We have then compared these predictions on tSZ with the Planck, SPT, and ACT data. Our findings are summarised as follows:
\begin{itemize}
\item The tSZ radial profile of Coma in the local universe
      simulation embedded in the background from the deep light-cone
      agrees well with that in the Planck data.
\item The local universe simulation predicts that the halo of Coma is moving away from us at $\approx400$ km/s with
      respect to the CMB rest frame, thus yielding a negative kSZ within
      the central region of Coma. The magnitude of kSZ is ten percent of
      tSZ at 150~GHz. On the other hand, because Coma is a merging
      system, we find a significant relative motion of the core (even
      increasing the negative signal in the center) and a significant
      positive kSZ in the outskirt, which comes from a infalling
      sub-structure moving toward us. This makes a positive kSZ
      contribution to tSZ at 150~GHz at distances beyond 1~Mpc from the
      center of Coma.
\item The predicted one-point PDF of the Compton $Y$ agrees
      with that measured by Planck, once the simulations are smoothed to
      the resolution of Planck's $Y$ map (10 arcmin). Given the much smaller beam size,
      we expect that ACT- and SPT-like instruments will see almost the full PDF that is resolved by
      our simulations. The tail of the full PDF follows a power-law
      with an index of $-3.2$.
\item The tSZ power spectrum measured from the simulation agrees with
      that of the Planck
      data at all multipoles up to $l\approx 1000$, once the power
      spectrum is rescaled to Planck 2015's ``TT+lowP+lensing''
      cosmological parameters with $\Omega_m=0.308$ and
      $\sigma_8=0.8149$ \citep{2015arXiv150201589P}. We have
      confirmed and understood this result using the analytical model.
\item Consistent with the previous work, we continue to find
      the predicted tSZ power spectrum at $l=3000$ that is significantly higher than that estimated by ACT and SPT. Whether this poses a challenge to theory is unclear, but our prediction is still well below the firm upper bound on tSZ given by the SPT data points with the primary CMB subtracted.
\item The simulation predicts the mean Compton $Y$ value of
      $1.18\times10^{-6}$ for $\Omega_m=0.272$ and
      $\sigma_8=0.809$. When the contributions from halos above a virial
      mass of $10^{13}~M_\odot/h$ are removed, we find $\bar
      Y=5\times10^{-7}$; thus, nearly half of the signal comes from such
      low-mass halos and diffuse gas outside halos. This remaining
      signal would pose a  challenge to detecting the primordial
      $y$-distortions.
\item Using the analytical model, we scale the Compton $Y$ value from the simulation to the Planck 2015 parameters with $\Omega_m=0.308$ and
      $\sigma_8=0.8149$, finding $\bar Y=1.57\times 10^{-6}$. This is
      still lower than, but not far away from, the new Planck bound, $\bar Y<2.2\times 10^{-6}$ \citep{2015arXiv150500781K}.
\item The one-point PDF and the power spectrum of kSZ from our
      simulations agree broadly with the previous work. While our box
      size is large, the contribution to kSZ is still dominated
by the largest modes within the box and originates mainly from high
      redshifts. Therefore, unlike for tSZ, we have not yet obtained a
      reliable, converged result on kSZ on large scales, $l\lesssim
      1000$. Simulations following even large cosmological volumes are needed.
\end{itemize}

In short, the main conclusion from our study is that all the
properties of tSZ found in the Magneticum Pathfinder simulation and the
local universe simulation agree well with the Planck
data. This includes the tSZ power spectrum, which was previously found to be in
tension with the Planck 2013 parameters
\citep{2014A&A...571A..21P,2014MNRAS.440.3645M}. Now, the tSZ
power spectrum calculated for the Planck
2015 parameters including CMB lensing information agrees with the
measurement at all multipoles up to $l\approx 1000$.


\section*{acknowledgments}

We thank N. Battaglia, J.~C. Hill, and A. Saro for many helpful
discussions, and R. Khatri for providing us with the PDFs of the Compton
$Y$ parameter estimated from the Planck data. E.K. thanks E. Jennings
for her help in calculating the mass function of Tinker et al. (2008).
K.D. acknowledges the support by the DFG Cluster
of Excellence ``Origin and Structure of the Universe''. We are especially grateful for
the support by M.~Petkova through the Computational Center for Particle and Astrophysics
(C$^2$PAP). Computations have been performed at the ``Leibniz-Rechenzentrum'' with CPU time 
assigned to the Project ``h0073''. Information on the {\it Magneticum
Pathfinder} project is available at http://www.magneticum.org.

\bibliographystyle{apj}

\bibliography{master,master3,Literaturdatenbank}
\label{lastpage}

\end{document}